\documentclass[lettersize,journal]{IEEEtran}
\usepackage{amsmath}
\usepackage{amssymb}
\DeclareMathOperator*{\argmax}{arg\,max}

\usepackage{graphicx}
\usepackage{epstopdf}
\usepackage{array}
\newcommand{\PreserveBackslash}[1]{\let\temp=\\#1\let\\=\temp}
\newcolumntype{C}[1]{>{\PreserveBackslash\centering}p{#1}}
\newcolumntype{R}[1]{>{\PreserveBackslash\raggedleft}p{#1}}
\newcolumntype{L}[1]{>{\PreserveBackslash\raggedright}p{#1}}
\usepackage[usenames]{color}
\usepackage{colortbl,booktabs}
\usepackage{tabularx}
\usepackage{multicol}
\usepackage{booktabs}
\usepackage[Symbol]{upgreek}
\usepackage{bm}
\usepackage{cite}
\usepackage{balance}
\usepackage{mathrsfs}
\usepackage{url}
\usepackage{threeparttable}
\usepackage{dcolumn}
\usepackage{multirow}
\usepackage{booktabs}
\usepackage{amsfonts}
\usepackage[linesnumbered,ruled]{algorithm2e}
\usepackage{float}
\usepackage{subfigure}
\usepackage{soul}
\usepackage[caption=false,font=normalsize,labelfont=sf,textfont=sf]{subfig}
\usepackage{textcomp}
\usepackage{stfloats}
\usepackage{verbatim}

\begin{document}
	\title{Beamforming Design with Partial Channel Estimation and Feedback for FDD RIS-Assisted Systems}

	\author{Xiaochun Ge, Shanping Yu, Wenqian Shen, \textsl{Member}, \textsl{IEEE}, Chengwen Xing, \textsl{Member}, \textsl{IEEE}, and Byonghyo Shim, \textsl{Senior Member}, \textsl{IEEE}

	\thanks{

		This work was supported by National Natural Science Foundation of China (Grant No. 72301030) and the NRF grant through the MSIT of Korea (Grant 2022R1A5A1027646).
		\textit{(Corresponding author: Shanping Yu.)}

		X. Ge, W. Shen, and C. Xing are with School of Information and Electronics, Beijing Institute of Technology, Beijing 100081, China (e-mail: xiaochun\_ge\_bit\_ee@163.com; shenwq@bit.edu.cn; xingchengwen@gmail.com).
	
		S.~Yu is with the School of Cyberspace Science and Technology, Beijing Institute of Technology, Beijing 100081, China (e-mail: ysp@bit.edu.cn).
		
		B. Shim is with the Institute of New Media and Communications, Department of Electrical and Computer Engineering, Seoul National University, Seoul 08826, South Korea (e-mail: bshim@snu.ac.kr).
	}
\vspace*{-5mm}				
}

	\maketitle
	\begin{abstract}
		Beamforming design with partial channel estimation and feedback for frequency-division duplexing (FDD) reconfigurable intelligent surface (RIS) assisted systems is considered in this paper.
		We leverage the observation that path angle information (PAI) varies more slowly than path gain information (PGI). Then, several dominant paths are selected among all the cascaded paths according to the known PAI for maximizing the spectral efficiency of downlink data transmission.
		To acquire the dominating path gain information (DPGI, also regarded as the path gains of selected dominant paths) at the base station (BS), we propose a DPGI estimation and feedback scheme by jointly beamforming design at BS and RIS.
		Both the required number of downlink pilot signals and the length of uplink feedback vector are reduced to the number of dominant paths, and thus we achieve a great reduction of the pilot overhead and feedback overhead.
		Furthermore, we optimize the active BS beamformer and passive RIS beamformer by exploiting the feedback DPGI to further improve the spectral efficiency.
		From numerical results, we demonstrate the superiority of our proposed algorithms over the conventional schemes.
	\end{abstract}
	
	\IEEEpeerreviewmaketitle
	
	\section{Introduction}\label{S1}
	
	\subsection{Motivation}

	Recently, reconfigurable intelligent surfaces (RISs) (also known as intelligent reflecting surfaces, IRSs) have been envisioned as a promising technique for the beyond fifth-generation (B5G) and sixth-generation (6G) wireless communication systems due to their potential to smartly reconfigure the wireless propagation environment in an energy-efficient and environment-friendly manner \cite{Background_RIS_LiuYuanwei_IEEECOMMS&T_2021,RIS_ModelExp_Dailinglong&Hanzo_Access_2020,RIS_ModelAny_Jinshi_JSAC_2020}.
	RISs are nearly-passive devices composed of arrays of reflecting elements which can reconfigure the incident signals \cite{Gong_RIS_1}.
	Specifically, if the channel state information (CSI) is perfectly known, the quality of wireless communication can be improved by adjusting the RIS reflection coefficients with the aid of a centralized controller \cite{Background_RIS_CuiShuguang_TWC_2020}.
	Therefore, the optimization of RIS reflection coefficients has been widely studied under different setups \cite{BF_RIS_HuangChongwen_TWC_2019,BF_RIS_JointA&P_ZhangRui_TWC_2019}, where the effectiveness of RIS in achieving high spectral efficiency with low energy and hardware cost is verified \cite{Background_RIS_CuiShuguang_TWC_2020,BF_RIS_Jinshi_TVT_2019,Gong_RIS_2}.

	To fully enjoy the potential benefits of RIS-assisted communication systems, the centralized controller needs to acquire the CSI accurately \cite{RIS_ChannelEst_Gaofeifei_TCOM_2022}.
	While most of the aforementioned literature assumes the perfect CSI is available \cite{CE_RIS_Tensor_Jinshi_TWC_2021}, in reality, the acquisition of CSI is by no means easy and very challenging.
	In the widely used time-division duplexing (TDD) systems, downlink CSI can be acquired by the uplink channel estimation according to the channel reciprocity between the uplink and downlink wireless channels \cite{Reciprocity_RIS_JinShi_IEEEWC_2021,FDDFB_RIS_DaiLinglong_GLOBECOM_2020,RIS_ModelExp_Jinshi_TWC_2021,RIS_TwotimeEst_Dailinglong_TCOM_2021}.$\footnote{The experimental results in \cite{Reciprocity_RIS_JinShi_IEEEWC_2021} validate that the channel reciprocity holds in RIS-assisted systems as long as the employed RISs are commonly designed and fabricated, and conform to the prerequisite of the Rayleigh-Carson reciprocity theorem, which has been discussed in detail in \cite{Reciprocity_RIS_JinShi_IEEEWC_2021}.}$
	Hence, existing works mainly consider the channel estimation problems in TDD RIS-assisted systems \cite{CE_RIS_ADMM_AiBo_Lett_2021,CE_RIS_YuWei_ArXiv_2019,CE_RIS_DaiLinglong_Lett_2021,RIS_ChannelEst_Gaofeifei_WCLett_2021,RIS_ChannelEst_Hanzo_TCOM_2021,CE_CS_WangPeilian_Lett_2020}.
	In practice, however, considering the difference between radio frequency (RF) circuits of the transmitting branch and the receiving branch, the required accuracy of antenna array calibration to maintain the channel reciprocity in TDD mode is extremely high \cite{TDD_Drawback_TWC_2016}.
	Therefore, it is of importance to come up with the design and optimization of RIS-assisted systems for the widely used frequency-division duplexing (FDD) mode, where the uplink and downlink channels are operated at different frequency bands \cite{FDDFB_RIS_DaiLinglong_TCOM_2021}.

	Since the channel reciprocity no longer holds in FDD systems, downlink CSI should be estimated using downlink pilot signals at the user equipment (UE) and then fed back to the base station (BS).
	However, the overhead of directly feeding back the downlink CSI is unaffordable in practice, especially for RIS-assisted systems with an extremely large number of RIS elements \cite{FDDFB_Adaptive_TWC_2021}.
	Although there is no path gain reciprocity between the uplink and downlink channels in FDD RIS-assisted systems, the angle reciprocity, a property that the angles of propagation paths are quite similar in the uplink and downlink channels, still holds \cite{FDDRIS_Tracking_TVT_2021}, which will be discussed subsequently.
	Therefore, in order to effectively reduce the feedback overhead in FDD RIS-assisted systems, we focus on estimating and feeding back the path gain information (PGI) \cite{FB_Pathsel_Shim_ICC_2019}, while the path angle information (PAI) can be obtained via the angle reciprocity \cite{FB_Pathsel_Shim_TWC_2020}.
	Furthermore, motivated by \cite{FB_Pathsel_Shim_ICC_2019}, we select several paths as dominant paths, and then estimate and feed back their corresponding dominating path gain information (DPGI) to further reduce pilot overhead and feedback overhead.

	\subsection{Related Work}

	Over the years, various channel estimation techniques for the RIS-assisted systems have been proposed \cite{CE_RIS_ADMM_AiBo_Lett_2021,CE_RIS_YuWei_ArXiv_2019,CE_RIS_DaiLinglong_Lett_2021,RIS_ChannelEst_Gaofeifei_WCLett_2021,RIS_ChannelEst_Hanzo_TCOM_2021,CE_CS_WangPeilian_Lett_2020}.
	In \cite{CE_RIS_ADMM_AiBo_Lett_2021} and \cite{CE_CS_WangPeilian_Lett_2020}, utilizing the sparse property of millimeter wave (mmWave) channels \cite{LowPS_GXC_TWC_2022,Xing_MM_MIMO},
	the compressive sensing (CS) based estimator and the alternating direction method of multipliers (ADMM) based estimator have been proposed, where both of which estimate the cascaded channel with a low training overhead.
	In \cite{CE_RIS_YuWei_ArXiv_2019}, a two-step channel estimation approach exploiting the common row-column-block sparsity structure among the uplink channel matrices of all users has been proposed.
	Although the aforementioned works \cite{CE_RIS_ADMM_AiBo_Lett_2021,CE_RIS_YuWei_ArXiv_2019,CE_RIS_DaiLinglong_Lett_2021,RIS_ChannelEst_Gaofeifei_WCLett_2021,RIS_ChannelEst_Hanzo_TCOM_2021,CE_CS_WangPeilian_Lett_2020} mainly consider the channel estimation problems for TDD mode, the estimators proposed can be easily extended or effectively applied to PAI acquisition at BS for FDD mode \cite{FDDFB_RIS_DaiLinglong_TCOM_2021}.

	Existing works focused on FDD RIS-assisted networks are relatively limited (see, e.g. \cite{FDDFB_RIS_DaiLinglong_GLOBECOM_2020,FDDFB_RIS_DaiLinglong_TCOM_2021,FDDFB_Adaptive_TWC_2021,FDDRIS_Tracking_TVT_2021,FDDRIS_ZhaoLian_ICC_2022,FDDRIS_Tao_WCNC_2021,FDDFB_RIS_Lusenbao_TVT_2022,FDD_RIS_Customization_Jinshi_TWC_2022}).
	Specifically, to avoid the performance degradation in practical application of RIS-assisted systems, authors in \cite{FDDRIS_Tracking_TVT_2021} and \cite{FDDRIS_ZhaoLian_ICC_2022} discussed the downlink channel tracking and the optimization of phase shifts at RIS, respectively.
	Moreover, a two-way passive RIS beamforming design has been proposed in \cite{FDDRIS_Tao_WCNC_2021}, where the passive beamformers for downlink and uplink are optimized simultaneously.
	Besides, authors in \cite{FDDFB_RIS_Lusenbao_TVT_2022} further extended the reflecting beamforming design to multi-user scenarios, which effectively shows the great potential of RIS in FDD systems.
	In addition, we note that the system performance of RIS-assisted FDD networks is often limited by the unaffordable CSI feedback overhead.
	In \cite{FDDFB_RIS_DaiLinglong_GLOBECOM_2020} and \cite{FDDFB_RIS_DaiLinglong_TCOM_2021}, the similarity among the RIS-UE channels of all users is exploited to reduce the CSI feedback overhead.
	The authors in \cite{FDDFB_Adaptive_TWC_2021} designed a cascaded codebook for the feedback of PGI by assuming that downlink CSI (including both PAI and PGI) is perfectly known at UE, and further carried out an in-depth study for multi-RIS-assisted systems in \cite{FDD_RIS_Customization_Jinshi_TWC_2022}.
	On this basis, we turn to consider the feedback of DPGI with a smaller dimension (rather than the whole PGI) by selecting several dominant paths to further reduce the feedback overhead, and the DPGI estimation scheme with low pilot overhead is also provided.

	\subsection{Main Contributions}

	In this paper, we propose a path selection technique with reduced pilot overhead and feedback overhead as well as a partial CSI-based beamforming design for the FDD RIS-assisted wireless communication systems.
	Our main contributions are summarized as follows:

	\begin{itemize}

	\item We propose a path selection strategy for the FDD RIS-assisted systems.
	It is observed that PAI varies more slowly than PGI, so PAI can be considered as unchanged and acquired by BS during a relatively long period called `angle coherence time' \cite{FB_MIMOAoD_ShenWenqian_TCOM_2018}.
	Hence, we calculate the contributions of all the cascaded paths to the spectral efficiency based on PAI, and sequentially remove the path with minimal contribution one by one, where the active and passive beamformers are optimized alternatively.
	Then, the remaining paths and their corresponding PGI are regarded as selected dominant paths and DPGI, respectively.
	In this way, the dimension of vector (corresponding to DPGI) to be estimated and fed back are effectively reduced without significant performance degradation.

	\item We propose a DPGI estimation and feedback scheme by exploiting PAI known at BS according to the angle reciprocity, where the accuracy of DPGI estimation is greatly improved by jointly designing the active BS beamforming and passive RIS beamforming.
	In our proposed scheme, we set the required number of downlink training pilot signals and the length of uplink feedback vector to the number of dominant paths.
	Therefore, we achieve a great reduction of the pilot and feedback overhead over the existing least square (LS)-based and minimum mean-square error (MMSE)-based estimators \cite{CE_CS_WangPeilian_Lett_2020}.$\footnote{The number of pilot signals required for existing LS- and MMSE-based estimators, which estimate the PGI with known PAI, is relatively large, often exceeding the number of cascaded paths \cite{Book_Statistical_Signal_Processing}.}$

	\item We propose an algorithm to alternatively update the active and passive beamformers on the basis of DPGI acquired at BS for further improving the spectral efficiency of downlink data transmission.
	From numerical results, we demonstrate the superiority of the proposed technique in terms of system performance, required pilot signals, and feedback overhead.

	\end{itemize}

	\subsection{Paper Outline}

	The rest of this paper is organized as follows. The system model of the RIS-assisted mmWave communications is presented in Section \ref{S2}.
	Then, the path selection technique, DPGI estimation and feedback scheme, and beamforming design are proposed in Section \ref{S3}.
	Simulation results are provided in Section \ref{S4}.
	Finally, we conclude our work in Section \ref{S5}.

	\subsection{Notation}

	In this paper, boldface lower-case and boldface capital letters represent column vectors and matrices, respectively.
	Besides, We denote
	$(\cdot)^*$, $(\cdot)^{\mathrm{T}}$, $(\cdot)^{\mathrm{H}}$, $|\cdot|$, $\|\cdot\|$, ${\|\cdot\|}_{\mathrm{F}}$, $(\cdot)^{-1}$, $\Re\{\cdot\}$, $\mathbb{E}\{\cdot\}$, and $\operatorname{Tr}\{\cdot\}$ as the conjugate, transpose, conjugate transpose, determinant of a matrix/absolute value of a scalar/cardinality of a set, Euclidean norm of a vector, Frobenius norm of a matrix, inverse, real part, statistical expectation, and trace operators, respectively.
	$\operatorname{vec}\{\cdot\}$ denotes the vectorization of a matrix (i.e., a linear transformation which stacks the columns of a matrix on top of one another to obtain a column vector), and $\operatorname{invec}\{\cdot\}$ denotes the inverse of vectorization.
	The operation $\arg (\mathbf{X})$ constructs a matrix by extracting the angles of all the elements in matrix $\mathbf{X}$.
	The Hadamard product and Kronecker product are denoted by $\odot$ and $\otimes$.
	In addition, all 0 matrix and all 1 matrix with dimension of $M \times N$ are represented by $\mathbf{0}_{M \times N}$, and $\mathbf{1}_{M \times N}$, respectively, and $\mathbf{I}_{M}$ denotes an identity matrix of size $M \times M$.
	$\mathbf{X}^{i,j}$ denotes the $\left(i,j\right)$-th element of matrix $\mathbf{X}$.
	Finally, $\mathcal{CN}(\mathbf{0}, \mathbf{R})$ denotes the zero-mean complex Gaussian distribution with covariance matrix $\mathbf{R}$.

	\section{System Model}\label{S2}

	In this section, we discuss the signal model of the RIS-assisted mmWave wireless communications and the angle reciprocity for the FDD RIS-assisted systems. We then explain the channel feedback mechanism for the FDD RIS-assisted wireless communication systems.

	\subsection{RIS-assisted mmWave Wireless Communication Model}\label{S2.1}

	In this paper, an RIS-assisted mmWave wireless communication system is investigated.
	Since the direct BS-UE channel has been studied extensively in many previous works and the blockage is a critical issue for mmWave wireless communications, we focus on the reflection link cascaded by the RIS \cite{FDDFB_RIS_DaiLinglong_TCOM_2021}.$\footnote{In scenarios where a direct link is present, the direct BS-UE channel is significantly stronger compared to the reflected channel \cite{Directpath_RIS_VPoor_TWC_2022,Directpath_RIS_VPoor_JSAC_2020}. In such cases, it becomes crucial to select the dominant paths from the direct link. Our work can be applied in this scenario with some simplifications, such as removing the passive RIS beamformer and making corresponding modifications.}$
	Through an RIS controller, BS can control the RIS to manipulate the electromagnetic response of incident waves.
	In our work, uniform planar arrays (UPAs) are deployed at both BS and RIS.
	By using the subscript `v' for vertical system parameters and the subscript `h' for horizontal system parameters, the number of BS antennas and RIS elements can be expressed as $N_{\mathrm{B}} = N_{\mathrm{B,v}} \times N_{\mathrm{B,h}}$ and $N_{\mathrm{R}} = N_{\mathrm{R,v}} \times N_{\mathrm{R,h}}$.
	The reflection coefficient matrix of RIS can be expressed as
	\begin{align}\label{Psi}
		\boldsymbol{\Psi} \! = \! \operatorname{diag}(\boldsymbol{\psi}) \! = \! \operatorname{diag} ( \! \left[ e^{j \psi_{1}} \! ,e^{j \psi_{2}} \! , \ldots, e^{j \psi_{N_{\mathrm{R}}}} \right]^{\rm{T}} \! ) \! \in \! \mathbb{C}^{N_{\mathrm{R}} \! \times \! N_{\mathrm{R}}}. \!
	\end{align}
	The cascaded channel vector $\mathbf{h} \in \mathbb{C}^{N_{\mathrm{B}} \times 1}$ between the BS and the single-antenna UE is
	\begin{align}\label{h_cascaded_inial}
		\mathbf{h}^{\rm{H}} = \mathbf{h}_{\mathrm{RU}}^{\rm{H}} \boldsymbol{\Psi} \mathbf{H}_{\mathrm{RB}},
	\end{align}
	where $\mathbf{H}_{\mathrm{RB}} \in \mathbb{C}^{N_{\mathrm{R}} \times N_{\mathrm{B}}}$ and $\mathbf{h}_{\mathrm{RU}} \in \mathbb{C}^{N_{\mathrm{R}} \times 1}$ denote the BS-RIS channel matrix and the RIS-UE channel vector, respectively.	
	According to the sparsity of mmWave channel, $\mathbf{H}_{\mathrm{RB}}$ can be expressed as the sum of several propagation paths, which is given by \cite{RIS_ChannelEst_Gaofeifei_TCOM_2022,LowPS_GXC_TWC_2022}
	\begin{align}\label{H_RB}
		\mathbf{H}_{\mathrm{RB}} = \sqrt{\frac{N_{\mathrm{B}} N_{\mathrm{R}}} {L_{\mathrm{RB}}}}  \mathbf{A}_{\mathrm{RB}} \operatorname{diag}(\boldsymbol{\alpha}) \mathbf{A}^{\rm{H}}_{\mathrm{B}} \in \mathbb{C}^{N_{\mathrm{R}} \times N_{\mathrm{B}}},
	\end{align}
	where $L_{\mathrm{RB}}$ and $\boldsymbol{\alpha} = \left[ \alpha_{1}, \alpha_{2}, \ldots, \alpha_{L_{\mathrm{RB}}} \right]^{\rm{T}} \in \mathbb{C}^{L_{\mathrm{RB}} \times 1}$ denote the number of paths in BS-RIS channel and the normalized complex gain with $\alpha_{p} \sim \mathcal{CN}(0,1)$ for $p = 1, 2, \ldots, L_{\mathrm{RB}}$, respectively.
	
	Similarly, $\mathbf{h}_{\mathrm{RU}}$ is given by	
	\begin{align}\label{h_RU}
		\mathbf{h}_{\mathrm{RU}} = \sqrt{\frac{N_{\mathrm{R}}}{L_{\mathrm{RU}}}} \mathbf{A}_{\mathrm{RU}} \boldsymbol{\beta}^{*} \in \mathbb{C}^{N_{\mathrm{R}} \times 1},
	\end{align}
	where $L_{\mathrm{RU}}$ and $\boldsymbol{\beta} = \left[\beta_{1}, \beta_{2}, \ldots, \beta_{L_{\mathrm{RU}}}\right]^{\rm{T}} \in \mathbb{C}^{L_{\mathrm{RU}} \times 1}$ are the number of paths in RIS-UE channel and the normalized complex gain with $\beta_{q} \sim \mathcal{C N}(0,1)$, for $q = 1, 2, \ldots, L_{\mathrm{RU}}$.
	Furthermore, $\mathbf{A}_{\mathrm{B}} = \left[ \mathbf{a}_{\mathrm{B}, 1}, \mathbf{a}_{\mathrm{B}, 2}, \ldots, \mathbf{a}_{\mathrm{B}, L_{\mathrm{RB}}} \right] \in \mathbb{C}^{N_{\mathrm{B}} \times L_{\mathrm{RB}}}$, $\mathbf{A}_{\mathrm{RB}} = \left[ \mathbf{a}_{\mathrm{RB}, 1}, \mathbf{a}_{\mathrm{RB}, 2}, \ldots, \mathbf{a}_{\mathrm{RB}, L_{\mathrm{RB}}} \right] \in \mathbb{C}^{N_{\mathrm{R}} \times L_{\mathrm{RB}}}$, and $\mathbf{A}_{\mathrm{RU}} = \left[\mathbf{a}_{\mathrm{RU}, 1}, \mathbf{a}_{\mathrm{RU}, 2}, \ldots, \mathbf{a}_{\mathrm{RU}, L_{\mathrm{RU}}}\right] \in \mathbb{C}^{N_{\mathrm{R}} \times L_{\mathrm{RU}}}$ are the transmitting array response matrix at BS, the receiving array response matrix at RIS and the transmitting array response matrix at RIS, respectively.
	For $p = 1, 2, \ldots, L_{\mathrm{RB}}$ and $q = 1, 2, \ldots, L_{\mathrm{RU}}$, we have
	\begin{align}
		& \mathbf{a}_{\mathrm{B}, p} = \mathbf{a}_{\mathrm{v}} \left( N_{\mathrm{B,v}}, \theta_{\mathrm{B,v}, p} \right) \otimes \mathbf{a}_{\mathrm{h}} \left( N_{\mathrm{B,h}}, \theta_{\mathrm{B,v}, p}, \theta_{\mathrm{B,h}, p} \right), \\
		& \mathbf{a}_{\mathrm{RB}, p} \! = \! \mathbf{a}_{\mathrm{v}} \! \left( N_{\mathrm{R,v}}, \phi_{\mathrm{RB,v}, p} \right) \! \otimes \! \mathbf{a}_{\mathrm{h}} \! \left( N_{\mathrm{R,h}}, \phi_{\mathrm{RB,v}, p}, \phi_{\mathrm{RB,h}, p} \right) \! , \! \\
		& \mathbf{a}_{\mathrm{RU}, q} \! = \! \mathbf{a}_{\mathrm{v}} \! \left( N_{\mathrm{R,v}}, \theta_{\mathrm{RU,v}, q} \right) \! \otimes \! \mathbf{a}_{\mathrm{h}} \! \left( N_{\mathrm{R,h}}, \theta_{\mathrm{RU,v}, q}, \theta_{\mathrm{RU,h}, q} \right) \! , \!
	\end{align}
	where the array response vectors of half-wavelength spaced UPAs are given by \cite{FDDFB_Adaptive_TWC_2021}
	\begin{align}
		\nonumber & \mathbf{a}_{\mathrm{v}} \left( N_{\mathrm{v}}, \theta_{\mathrm{v}} \right) = \sqrt{\frac{1} {N_{\mathrm{v}}}} \left[ 1, e^{j \pi \cos \left(\theta_{\mathrm{v}}\right)}, \ldots, \right. \\
		& \qquad \qquad \qquad \left. e^{j \pi \left(N_{\mathrm{v}}-1 \right) \cos \left(\theta_{\mathrm{v}}\right)} \right]^{\rm{T}} \in \mathbb{C} ^{N_{\mathrm{v}} \times 1}, \\
		\nonumber & \mathbf{a}_{\mathrm{h}} \left( N_{\mathrm{h}}, \theta_{\mathrm{v}}, \theta_{\mathrm{h}} \right) = \sqrt{\frac{1} {N_{\mathrm{h}}}} \left[ 1, e^{j \pi \sin \left(\theta_{\mathrm{v}}\right) \sin \left(\theta_{\mathrm{h}}\right)}, \ldots, \right. \\
		& \qquad \qquad \qquad \quad \left. e^{j \pi \left(N_{\mathrm{h}}-1\right) \sin \left(\theta_{\mathrm{v}}\right) \sin \left(\theta_{\mathrm{h}}\right)} \right]^{\rm{T}}  \in \mathbb{C} ^{N_{\mathrm{h}} \times 1}
	\end{align}
	with $\theta_{\mathrm{B,v}, p}$, $\phi_{\mathrm{RB,v}, p}$, and $\theta_{\mathrm{RU,v}, q}$ ($\theta_{\mathrm{B,h}, p}$, $\phi_{\mathrm{RB,h}, p}$, and $\theta_{\mathrm{RU,h}, q}$) denoting the angle of departure (AoD) of the $p$-th path for BS-RIS channel $\mathbf{H}_{\mathrm{RB}}$, the angle of arrival (AoA) of the $p$-th path for BS-RIS channel $\mathbf{H}_{\mathrm{RB}}$ and the AoD of the $q$-th path for RIS-UE channel $\mathbf{h}_{\mathrm{RU}}$ in the vertical (horizontal) direction, respectively.
	Then, the channel vector in (\ref{h_cascaded_inial}) can be rewritten  as
	\begin{align}\label{h_cascaded}
		\mathbf{h}^{\rm{H}} = \sqrt{\frac{N_{\mathrm{B}} N_{\mathrm{R}}^{2}}{L}} \boldsymbol{\beta}^{\rm{T}} \mathbf{A}_{\mathrm{RU}}^{\rm{H}} \boldsymbol{\Psi} \mathbf{A}_{\mathrm{RB}} \operatorname{diag}(\boldsymbol{\alpha}) \mathbf{A}_{\mathrm{B}}^{\rm{H}},
	\end{align}	
	where $L = L_{\mathrm{RB}} L_{\mathrm{RU}}$ represents the total number of cascaded paths.

	For the RIS-assisted wireless communication model, the downlink signal received at UE can be expressed as
	\begin{align}\label{y_dt}
		y = \sqrt{P_\mathrm{t}} \mathbf{h}^{\rm{H}} \mathbf{f}_{\mathrm{t}} s + n,
	\end{align}
	where $P_\mathrm{t}$ is the transmitting power,
	$\mathbf{f}_{\mathrm{t}} \in \mathbb{C}^{N_{\mathrm{B}} \times 1}$ is the active beamformer at BS,
	$s$ is the signal transmitted from BS satisfying $\mathbb{E}\left[s s^{*}\right]=1$, and $n \sim \mathcal{C N}\left(0, \sigma_{n}^{2}\right)$ is the complex Gaussian noise with noise power $\sigma_{n}^{2}$.
	Then, the achievable downlink spectral efficiency $R$ is \cite{FB_Pathsel_Shim_ICC_2019,Xing_Matrix_Monotonic_2}
	\begin{align}\label{R_appro}
		R = \log_{2}\left(1+\frac{P_{\mathrm{t}}}{\sigma_{n}^{2}} \mathbb{E}\left[\left|\mathbf{h}^{\rm{H}} \mathbf{f}_{\mathrm{t}}\right|^{2}\right]\right).
	\end{align}

	\subsection{Angle Reciprocity for FDD RIS-assisted Systems}\label{S2.2}

	It is observed that only the signal components which physically reverse uplink propagation paths can be transmitted in the downlink for FDD communication systems \cite{FB_Pathsel_Shim_ICC_2019}.	
	Hence, when the carrier frequencies between downlink and uplink channels do not differ too much (typically less than a few GHz), although their PGIs differ from each other, uplink PAI and downlink PAI are fairly similar.
	This phenomenon is referred to as the angle reciprocity \cite{FB_Pathsel_Shim_ICC_2019}.
	In addition, the introduction of commonly designed and fabricated RISs does not impair the angle reciprocity between the uplink and downlink channels, which has been demonstrated through experimental results under different conditions
	\cite{Reciprocity_RIS_JinShi_IEEEWC_2021,Reciprocity_RIS_JinShi_TAP_2022}. Therefore, BS can estimate the uplink PAI using the pilot signals sent from UE, and then exploit the estimated PAI for the downlink beamforming design according to the angle reciprocity in FDD RIS-assisted systems \cite{FDDRIS_Tracking_TVT_2021}.
	In addition, channel estimation problems for the RIS-assisted systems have been widely studied (see, e.g. \cite{CE_RIS_YuWei_ArXiv_2019,CE_RIS_ADMM_AiBo_Lett_2021,CE_RIS_DaiLinglong_Lett_2021,CE_CS_WangPeilian_Lett_2020}). Although many of them consider the TDD mode, the estimators can be easily extended or effectively applied to the acquisition of PAI at BS for the FDD mode \cite{FDDFB_RIS_DaiLinglong_TCOM_2021,Xing_Wideband_MIMO}.
	For example, the PAI (including AoAs and AoDs) is quantized into the two-dimensional discrete angular grids in \cite{CE_CS_WangPeilian_Lett_2020}, and then the CS-based technique is used to estimate the positions of non-zero elements in the grids \cite{CE_CS_WangPeilian_Lett_2020}.
	
	\subsection{Channel Feedback for FDD RIS-assisted Systems}\label{S2.3}

	In the  FDD systems, the downlink CSI fed back from UE is essential for the beamforming design.
	The random vector quantization (RVQ) codebook is widely used for the CSI feedback, which is randomly generated by selecting vectors independently from the uniform distribution on the complex unit sphere \cite{FB_MIMOAoD_ShenWenqian_TCOM_2018}.
	In this scheme, UE first normalizes the vector $\mathbf{z} \in \mathbb{C}^{N \times 1}$ to be fed back as $\overline{\mathbf{z}} = \frac{\mathbf{z}}{\left\|\mathbf{z}\right\|}$, and then feeds back the codeword $\hat{b}$ satisfying \cite{FB_MIMOAoD_ShenWenqian_TCOM_2018}
 	\begin{align}\label{RVQ_defi}
		\hat{b} = \argmax_{b \in\left\{1,2, \cdots, 2^B \right\}} \left|\bar{\mathbf{z}}^{\rm{H}} \mathbf{c}_b\right|^2,
	\end{align}	
	where $\mathbf{C}_{\mathrm{RVQ}}=\left[\mathbf{c}_{1}, \mathbf{c}_{2}, \ldots, \mathbf{c}_{2^{B}}\right] \in \mathbb{C}^{N \times 2^{B}}$ is the pre-defined $B$-bits RVQ codebook with $\left\|\mathbf{c}_{b}\right\|^{2}=1$ for $b=1,2, \ldots, 2^{B}$.$\footnote{We follow the common assumption in channel feedback that the scalar $\left\|\mathbf{z}\right\|$ (i.e., the magnitude of $\mathbf{z}$) can be fed back perfectly, thereby the more challenging feedback of the vector $\bar{\mathbf{z}}$ (i.e., the direction of $\mathbf{z}$) is focused in this study \cite{FB_MIMOAoD_ShenWenqian_TCOM_2018}.}$
	To properly control the quantization distortion, the required number of feedback bits is given by $B \approx \frac{(N-1)}{3} \times \mathrm{SNR}$, where $\mathrm{SNR}$ denotes the signal-to-noise-ratio for codeword transmission \cite{FB_Pathsel_Shim_ICC_2019}.
	However, for a FDD RIS-assisted communication system with $N_{\mathrm{B}}$ antennas at BS and $N_{\mathrm{R}}$ reflecting elements at RIS, the overhead of directly feeding back downlink CSI (including both PAI and PGI) with dimension of $N_{\mathrm{B}} N_{\mathrm{R}} \times 1$ is unbearable, which leads to an extremely huge number of feedback bits to achieve an acceptable feedback distortion \cite{FDDFB_Adaptive_TWC_2021}.
	Fortunately, downlink PAI can be acquired by the uplink channel estimation via the angle reciprocity and the slowly-varying characteristic of PAI.
	Then, only downlink PGI with dimension of $L \times 1$ needs to be fed back.
	Thus, the dimension of feedback vector can be reduced from $N_{\mathrm{B}} N_{\mathrm{R}} \times 1$ to $L \times 1$, where $L = L_{\mathrm{RB}} L_{\mathrm{RU}}$ represents the total number of cascaded paths \cite{FDDFB_Adaptive_TWC_2021}.
	The purpose of our work is to further reduce the feedback overhead in FDD RIS-assisted systems.
	Inspired by \cite{FB_Pathsel_Shim_ICC_2019}, we choose a few dominant paths maximizing the spectral efficiency from all the cascaded paths, and feed back corresponding DPGI instead of the overall PGI.

	\section{Proposed Path Selection Based Scheme}\label{S3}	

\begin{figure*}[t]
		\center{\includegraphics[width=0.8\textwidth]{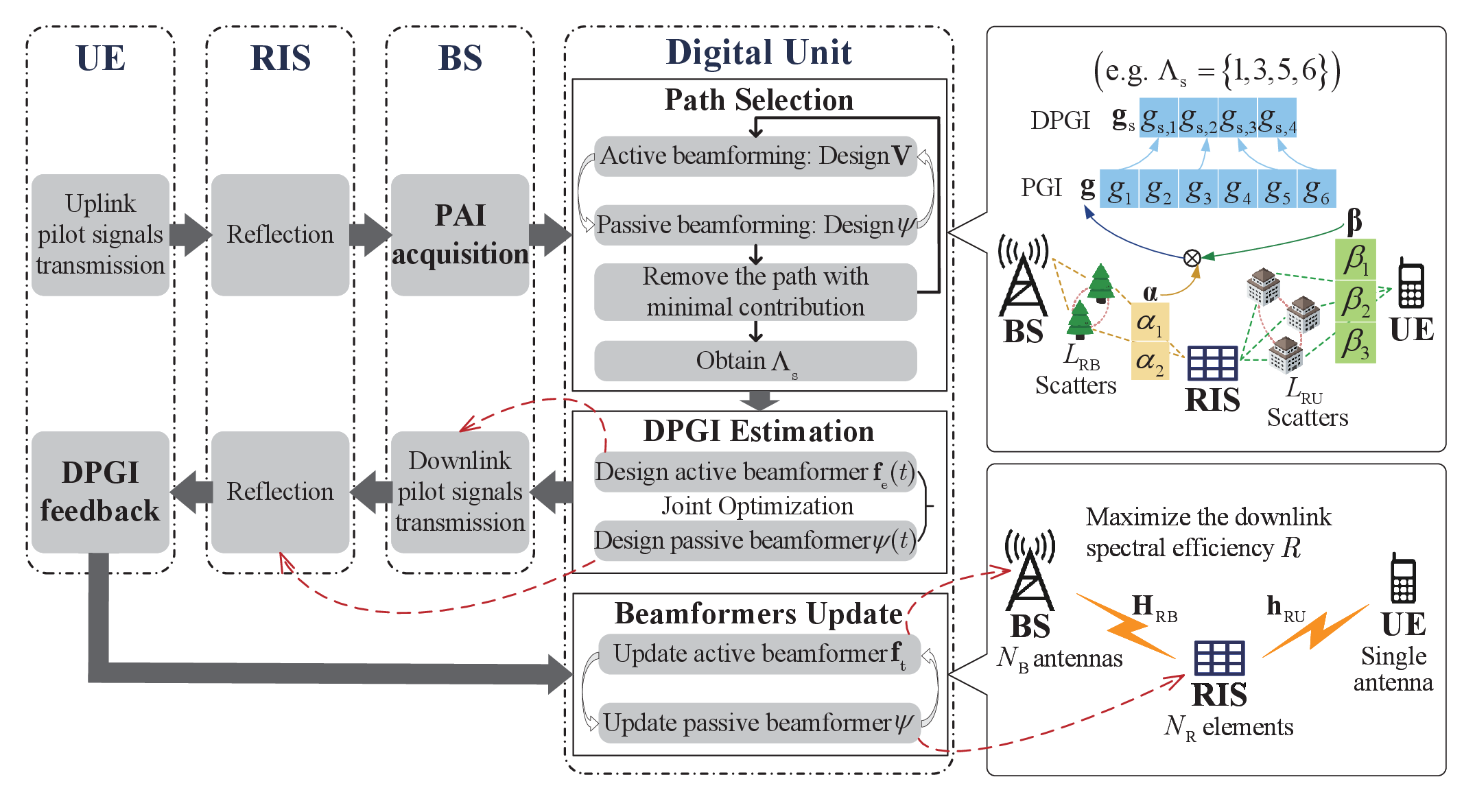}}
		\caption{Overall strategy and steps of the proposed path selection based feedback reduction and beamforming design scheme.}
		\label{fig1}
	\end{figure*}

	In this section, we present the path selection based feedback reduction and beamforming design scheme.
	The overall strategy and main steps are summarized in Fig.~\ref{fig1}.

	\begin{itemize}
		\item \textit{Step 1 (PAI acquisition):} BS estimates the uplink PAI using the pilot signals sent from UE and reflected by RIS (the feasibility of this step has been discussed in the aforementioned section, which will not be repeated in the following).$\footnote{The cost of PAI estimation is relatively small, primarily due to two reasons: first, the utilization of CS-based sparse recovery algorithms reduces the pilot overhead according to the sparsity of mmWave channels \cite{CE_CS_WangPeilian_Lett_2020}; second, the PAI estimation only needs to be performed once within the angle coherence time \cite{FB_MIMOAoD_ShenWenqian_TCOM_2018}.}$
		According to the angle reciprocity of FDD systems, BS obtains the downlink PAI by reversing the uplink PAI \cite{FDDRIS_Tracking_TVT_2021}.
		\item \textit{Step 2 (Path selection):} BS selects several dominant paths according to the proposed path selection strategy, during which alternating optimization of active and passive beamformers is performed to make sure that the selected dominant paths contribute to the maximization of downlink spectral efficiency $R$.
		\item \textit{Step 3 (DPGI estimation and feedback):} Downlink pilot signals are transmitted at BS and reflected by RIS with jointly designed active and passive beamforming vectors for DPGI estimation. By performing the proposed estimation and feedback scheme at UE, BS obtains the quantized downlink DPGI through uplink feedback.
		\item \textit{Step 4 (Beamformers update):} The active and passive beamformers for downlink data transmission will be updated alternatively based on the fed back DPGI to further improve the spectral efficiency $R$.
	\end{itemize}
	The corresponding steps involved will be discussed in detail as follows.

	\subsection{Path Selection}\label{S3.1}

	In this subsection, we discuss the proposed dominating path selection strategy.

	\textit{1) Problem Formulation:}

	According to (\ref{R_appro}), the spectral efficiency $R$ can be improved by maximizing $\mathbb{E} \left[\left|\mathbf{h}^{\rm{H}} \mathbf{f}_{\mathrm{t}}\right|^{2} \right]$.
	Thus, we begin with rewriting the cascaded channel vector $\mathbf{h} \in \mathbb{C}^{N_{\mathrm{B}} \times 1}$ in (\ref{h_cascaded}) as
	\begin{align}\label{h_Ag}
		\mathbf{h} = \mathbf{A} \mathbf{g}^{*},
	\end{align}
	where $\mathbf{g} = \boldsymbol{\beta} \otimes \boldsymbol{\alpha} \in \mathbb{C}^{L \times 1}$ is the cascaded PGI.
	The matrix $\mathbf{A}$ can be expressed as
	\begin{align}
		& \mathbf{A} = \left[\mathbf{a}_{1}, \mathbf{a}_{2}, \ldots, \mathbf{a}_{L}\right] \in \mathbb{C}^{N_{\mathrm{B}} \times L}, \label{A_defi_col} \\
		& \mathbf{a}_{l}=\mathbf{B}_{l} \boldsymbol{\psi}^{*} \in \mathbb{ C}^{N_{\mathrm{B}} \times 1}, \label{B_defi_1} \\ 
		& \mathbf{B}_{l} = \sqrt{\frac{N_{\mathrm{B}} N_{\mathrm{R}}^{2}}{L}} \mathbf{a}_{\mathrm{B}, p} \mathbf{a}_{\mathrm{RB}, p}^{\rm{H}} \operatorname{diag}\left(\mathbf{a}_{\mathrm{RU}, q}\right) \in \mathbb{C}^{N_{\mathrm{B}} \times N_{\mathrm{R}}} \label{B_defi_2}
	\end{align}
	for $l = 1, 2, \ldots, L$.
	The index $l$ of cascaded paths is given by $l = (q-1) L_{\mathrm{RB}} + p$ for $p = 1, 2, \ldots, L_{\mathrm{RB}}$ and $q = 1, 2, \ldots, L_{\mathrm{RU}}$.

	Initially, all the $L$ paths are in the selected set.
	Then, the path with minimal contribution to $R$ is removed sequentially until the number of remaining paths reaches the pre-defined number of dominant paths.
	In this iterative process, we denote $L_\mathrm{s}$ and $\Lambda_{\mathrm{s}} \subseteq\{1,2, \ldots, L\}$ as the number of remaining dominant paths and the set of corresponding indices satisfying $\left| \Lambda_{\mathrm{s}} \right| = L_{\mathrm{s}}$.
	Similarly, the number of removed paths and the set of corresponding indices are denoted as $L_{\mathrm{r}}$ and $\Lambda_{\mathrm{r}} \subseteq\{1,2, \ldots, L\}$, where $L_{\mathrm{r}}= L - L_{\mathrm{s}}$, $\left| \Lambda_{\mathrm{r}} \right| = L_{\mathrm{r}}$.
	Thus, the cascaded channel in (\ref{h_Ag}) can be decomposed as
	\begin{align}\label{h_AsArg}
		\mathbf{h} = \mathbf{A}_{\mathrm{s}} \mathbf{g}^{*}_{\mathrm{s}}+\mathbf{A}_{\mathrm{r}} \mathbf{g}^{*}_{\mathrm{r} },
	\end{align}
	where $\mathbf{g}_{\mathrm{s}} = \left[ g_{\mathrm{s},1}, g_{\mathrm{s},2}, \ldots, g_{\mathrm{s},L_{\mathrm{s}}} \right]^{\rm{T}} \in \mathbb{C}^{L_{\mathrm{s}} \times 1}$ and $\mathbf{g}_{\mathrm{r}} = \left[ g_{\mathrm{r},1}, g_{\mathrm{r},2}, \ldots, g_{\mathrm{r},L_{\mathrm{r}}} \right]^{\rm{T}} \in \mathbb{C}^{L_{\mathrm{r}} \times 1}$ are the vectors composed of the corresponding elements in $\mathbf{g}$ according to $\Lambda_{\mathrm{s}}$ and $\Lambda_{\mathrm{r}}$, and $\mathbf{g}_{\mathrm{s}} \in \mathbb{C}^{L_{\mathrm{s}} \times 1}$ is denoted as the so-called DPGI in the following.
	Similarly, according to $\Lambda_{\mathrm{s}}$ and $\Lambda_{\mathrm{r}}$,  the sub-matrices composed of the corresponding column vectors in $\mathbf{A}$ are expressed as $\mathbf{A}_{\mathrm{s}} = \left[ \mathbf{a}_{\mathrm{s},1}, \mathbf{a}_{\mathrm{s},2}, \ldots, \mathbf{a}_{\mathrm{s},L_{\mathrm{s}}} \right] \in \mathbb{C}^{N_{\mathrm{B}} \times L_{\mathrm{s}}}$ and $\mathbf{A}_{\mathrm{r}} = \left[ \mathbf{a}_{\mathrm{r},1}, \mathbf{a}_{\mathrm{r},2}, \ldots, \mathbf{a}_{\mathrm{r},L_{\mathrm{r}}} \right] \in \mathbb{C}^{N_{\mathrm{B}} \times L_{\mathrm{r}}}$.
	Here we decompose the active beamformer $\mathbf{f}_{\mathrm{t}} \in \mathbb{C}^{N_{\mathrm{B}} \times 1}$ into the product of two parts \cite{FB_Pathsel_Shim_ICC_2019,FB_Pathsel_Shim_TWC_2020}:
	\begin{align}\label{f_Vg}
		\mathbf{f}_{\mathrm{t}} = \mathbf{V} \mathbf{g}^{*}_{\mathrm{s}},
	\end{align}
	where $\mathbf{V} \in \mathbb{C}^{N_{\mathrm{B}} \times L_\mathrm{s}}$ is the active beamforming matrix to be optimized.	
	
	\noindent \textbf{Lemma 1:}	
	By substituting (\ref{h_AsArg}) and (\ref{f_Vg}) into the objective function $\mathbb{E}\left[\left|\mathbf{h}^{\rm{H}} \mathbf{f}_{\mathrm{t}}\right|^{2}\right]$, we have
	\begin{align}\label{goal_inial}
		\mathbb{E}\left[\left|\mathbf{h}^{\rm{H}} \mathbf{f}_{\mathrm{t}}\right|^{2}\right] = \ & 
		\nonumber (4 - 2 Q) \left\|\operatorname{diag} \left(\mathbf{A}^{\rm{H}}_{\mathrm{s}} \mathbf{V}\right)\right\|^{2} + Q \left|\operatorname{tr}(\mathbf{A}^{\rm{H}}_{\mathrm{s}} \mathbf{V})\right|^{2} \\
		& + Q \left\| \mathbf{A}^{\rm{H}}_{\mathrm{s}} \mathbf{V} \right\|_{\mathrm{F}}^{2} + Q \left\| \mathbf{A}^{\rm{H}}_{\mathrm{r}} \mathbf{V} \right\|^{2}_{\rm{F}},
	\end{align}
	where $Q = \frac{L + L_{\mathrm{RB}} + L_{\mathrm{RU}} - 3}{L - 1}$ is defined for notational simplicity.
	\begin{IEEEproof}[Proof]
		See Appendix A.
	\end{IEEEproof}

	In the process of path selection, the index of the path with minimal contribution to $R$ is removed from $\{\Lambda_{\mathrm{s}}\}$ sequentially, and the active and passive beamformers need to be optimized. Based on the above derivations, the initial optimization problem is expressed as
	\begin{align}\label{Opt_inial}
		\mathcal{P}_1 \! : \! \max_{\{\Lambda_{\mathrm{s}}, \mathbf{V}, \boldsymbol{\psi}\}} \! R, 
		\ {\text{s.t.}} \left| \Lambda_{\mathrm{s}} \right| \! = \! L_{\mathrm{s}}, 
		\left\| \! \mathbf{V} \mathbf{g}^{*}_{\mathrm{s}}\right\|^{2} \!\! = \! 1, 
		\left|\boldsymbol{\psi}\right| \! = \! \mathbf{1}_{N_{\rm{R}} \times 1}, \!
	\end{align}	
	where the second constraint in (\ref{Opt_inial}) is due to the power constraint $\left\|\mathbf{f}_{\mathrm{t}}\right\|^{2} = \left\|\mathbf{V}\mathbf{g}^{*}_{\mathrm{s}}\right\|^{2} = 1$.
	For a given $\Lambda_{\mathrm{s}}$, the initial optimization problem $\mathcal{P}_1$ can be simplified as
	\begin{align}\label{Opt_1}
		\mathcal{P}_2: \ \max_{\{\mathbf{V}, \boldsymbol{\psi}\}} \mathbb{E}\left[\left|\mathbf{h}^{\rm{H}} \mathbf{f}_{\mathrm{t}}\right|^{2}\right] \! , 
		\ {\text{s.t.}} \ \left\|\mathbf{V}\right\|_{\rm{F}}^{2} \! = \! 1, 
		\left|\boldsymbol{\psi}\right| \! = \! \mathbf{1}_{N_{\rm{R}} \times 1}, \!
	\end{align}	
	where the first constraint in (\ref{Opt_1}) is adopted instead of the second constraint in (\ref{Opt_inial}) because $\mathbf{g}_{\mathrm{s}}$ (DPGI) has not been acquired at BS during the process of path selection.$\footnote{We note that this approximation scales the contributions to $\mathbb{E}\left[\left|\mathbf{h}^{\rm{H}} \mathbf{f}_{\mathrm{t}}\right|^{2}\right]$ (the components related to each path in $\mathbb{E}\left[\left|\mathbf{h}^{\rm{H}} \mathbf{f}_{\mathrm{t}}\right|^{2}\right]$ are separated and formulated in (\ref{Path Contribution}) as below) of all the remaining dominant paths with the same proportion $\frac{\left\|\mathbf{V} \mathbf{g}^{*}_{\mathrm{s}}\right\|^2} {\left\|\mathbf{V}\right\|_{\rm{F}}^2}$, so it will not result in performance degradation for the path selection.}$
	Due to the coupling of $\bf{V}$ and $\boldsymbol{\psi}$ in the objective function $\mathbb{E}\left[\left|\mathbf{h}^{\rm{H}} \mathbf{f}_{\mathrm{t}}\right|^{2}\right]$ of (\ref{Opt_1}), they cannot be optimized jointly, and thus we employ the alternating optimization of the active and passive beamformers.

	\textit{2) Active Beamforming Design:}

	We first consider the design of active beamforming matrix $\bf{V}$ with fixed $\Lambda_{\mathrm{s}}$ and $\boldsymbol{\psi}$, which can be given by
	\begin{align}
		\mathcal{P}_3: \ \max_{\{\mathbf{V}\}} \mathbb{E}\left[\left|\mathbf{h}^{\rm{H}} \mathbf{f}_{\mathrm{t}}\right|^{2}\right], 
		\ {\text{s.t.}} \ \left\|\mathbf{V}\right\|_{\rm{F}}^{2} = 1 \label{Opt_2}.
	\end{align}	
	On the basis of (\ref{goal_inial}), the objective function of (\ref{Opt_2}) can be equivalently rewritten as
	\begin{align}\label{objective_P_3}
		\mathbb{E}\left[\left|\mathbf{h}^{\rm{H}} \mathbf{f}_{\mathrm{t}}\right|^{2}\right]
		= \mathbf{v}^{\rm{H}}{\mathbf{J}_\mathrm{act}}\mathbf{v},
	\end{align}	
	where $\mathbf{v} = \operatorname{vec}(\mathbf{V}) \in \mathbb{C}^{N_\mathrm{B} L_{\mathrm{s}} \times 1}$ is the vectorization of active beamforming matrix $\mathbf{V}$, and $\nonumber {\mathbf{J}_\mathrm{act}} \in \mathbb{C}^{N_{\mathrm{B}} L_{\mathrm{s}} \times N_{\mathrm{B}} L_{\mathrm{s}}}$ is
	\begin{align}\label{J_act}
		\nonumber {\mathbf{J}_\mathrm{act}} = 
		& \ (4 - 2 Q) \operatorname{diag}\left( \mathbf{a}_{\mathrm{s}, 1}\mathbf{a}_{\mathrm{s}, 1}^{\rm{H}}, \mathbf{a}_{\mathrm{s}, 2}\mathbf{a}_{\mathrm{s}, 2}^{\rm{H}}, \ldots, \mathbf{a}_{\mathrm{s}, L_\mathrm{s}}\mathbf{a}_{\mathrm{s}, L_\mathrm{s}}^{\rm{H}}\right) \\
		\nonumber & + Q \operatorname{vec}\left(\mathbf{A}_{\mathrm{s}}\right) \! \left(\operatorname{vec} \left(\mathbf{A}_{\mathrm{s}}\right)\right)^{\rm{H}}
		\! + \! Q \left(\mathbf{I}_{L_{\mathrm{s}}} \! \otimes \! \mathbf{A}_{\mathrm{s}}^{\rm{H}}\right)^{\rm{H}} \! \left(\mathbf{I}_{L_{\mathrm{s}}} \! \otimes \! \mathbf{A}_{\mathrm{s}}^{\rm{H}}\right) \\
		& + Q \left(\mathbf{I}_{L_{\mathrm{s}}} \otimes \mathbf{A}_{\mathrm{r}}^{\rm{H}}\right)^{\rm{H}}\left(\mathbf{I}_{L_{\mathrm{s}}} \otimes \mathbf{A}_{\mathrm{r}}^{\rm{H}}\right),
	\end{align}
	where the following properties are used
	\begin{align}
		& \left\|\operatorname{diag} \left(\mathbf{A}^{\mathrm{H}}_{\mathrm{s}} \mathbf{V}\right)\right\|^{2} \! = \! \mathbf{v}^{\mathrm{H}} \! \operatorname{diag}\left( \mathbf{a}_{\mathrm{s}, 1}\mathbf{a}_{\mathrm{s}, 1}^{\mathrm{H}}, \ldots, \mathbf{a}_{\mathrm{s}, L_\mathrm{s}}\mathbf{a}_{\mathrm{s}, L_\mathrm{s}}^{\mathrm{H}}\right) \! \mathbf{v}, \! \\
		& \left|\operatorname{tr}(\mathbf{A}^{\mathrm{H}}_{\mathrm{s}} \mathbf{V})\right|^{2} = \mathbf{v}^{\mathrm{H}} \operatorname{vec}\left(\mathbf{A}_{\mathrm{s}}\right)\left(\operatorname{vec}\left(\mathbf{A}_{\mathrm{s}}\right)\right)^{\mathrm{H}} \mathbf{v}, \\
		& \left\|\mathbf{A}_{\mathrm{s}}^{\mathrm{H}} \mathbf{V} \right\|_{\mathrm{F}}^{2} = \mathbf{v}^{\mathrm{H}} \left(\mathbf{I}_{L_{\mathrm{s}}} \otimes \mathbf{A}_{\mathrm{s}}^{\mathrm{H}} \right)^{\mathrm{H}} \left(\mathbf{I}_{L_{\mathrm{s}}} \otimes \mathbf{A}_{\mathrm{s}}^{\mathrm{H}} \right) \mathbf{v}, \\
		& \left\|\mathbf{A}_{\mathrm{r}}^{\mathrm{H}} \mathbf{V} \right\|_{\mathrm{F}}^{2} = \mathbf{v}^{\mathrm{H}} \left(\mathbf{I}_{L_{\mathrm{s}}} \otimes \mathbf{A}_{\mathrm{r}}^{\mathrm{H}} \right)^{\mathrm{H}} \left(\mathbf{I}_{L_{\mathrm{s}}} \otimes \mathbf{A}_{\mathrm{r}}^{\mathrm{H}} \right) \mathbf{v}.
	\end{align}
	Therefore, based on the formulation of (\ref{objective_P_3}), the optimal solution $\bf{V}^{\star}$ of $\mathcal{P}_3$ is given by \cite{FB_Pathsel_Shim_ICC_2019}
	\begin{align}
		\mathbf{V}^{\star} = \frac{1}{\left\|{\mathbf{u}_{\mathrm{act}, max}} \right\|} \operatorname{invec}\left({\mathbf{u}_{\mathrm{act}, max}} \right) \in \mathbb{C}^{N_{\mathrm{B}} \times L_{\mathrm{s}}},
	\end{align}
	where $\mathbf{u}_{\mathrm{act}, max} \in \mathbb{C}^{N_{\mathrm{B}} L_{\mathrm{s}} \times 1}$ is the eigenvector corresponding to the largest eigenvalue of ${\mathbf{J}_\mathrm{act}}$.

	\textit{3) Passive Beamforming Design:}

	After the optimization of the active beamforming matrix $\mathbf{V}$, we design the passive beamforming vector $\boldsymbol{\psi}$ with fixed $\Lambda_{\mathrm{s}}$ and $\mathbf{V}$. The optimization problem is formulated as
	\begin{align}\label{Opt_3}
		\mathcal{P}_4: \ \max_{\{\boldsymbol{\psi}\}} \mathbb{E}\left[\left|\mathbf{h}^{\rm{H}} \mathbf{f}_{\mathrm{t}}\right|^{2}\right],
		\ {\text{s.t.}} \ \left|\boldsymbol{\psi}\right| = \mathbf{1}_{N_{\rm{R}} \times 1}.
	\end{align}		
	The passive beamformer $\boldsymbol{\psi}$ cannot be directly optimized because it is an implicit variable embedded within matrices $\mathbf{A}_{\mathrm{s}}$ and $\mathbf{A}_{\mathrm{r}}$ in (\ref{goal_inial}).
	Therefore, by introducing a series of intermediate variables and performing matrix transformations, we will extract the passive beamformer $\boldsymbol{\psi}$ as follows.
	According to (\ref{B_defi_1}) and (\ref{B_defi_2}),  by defining $\mathbf{B}_{\mathrm{s}, l_{\mathrm{s}}} \in \mathbb{C}^{N_{\mathrm{B}} \times N_{\mathrm{R}}}$ and $\mathbf{B}_{\mathrm{r}, l_{\mathrm{r}}} \in \mathbb{C}^{N_{\mathrm{B}} \times N_{\mathrm{R}}}$ corresponding to the selected $L_{\mathrm{s}}$ dominant paths and the removed $L_{\mathrm{r}}$ paths, respectively,  we have
	\begin{align}
		& \mathbf{a}_{\mathrm{s}, l_{\mathrm{s}}} = \mathbf{B}_{\mathrm{s}, l_{\mathrm{s}}} \boldsymbol{\psi}^{*} \in \mathbb{C}^{N_{\mathrm{B}} \times 1} \ \text{for} \ l_{\mathrm{s}} = 1,2,\ldots, L_\mathrm{s}, \\
		& \mathbf{a}_{\mathrm{r}, l_{\mathrm{r}}} = \mathbf{B}_{\mathrm{r}, l_{\mathrm{r}}} \boldsymbol{\psi}^{*} \in \mathbb{C}^{N_{\mathrm{B}} \times 1}\  \text{for} \ l_{\mathrm{r}} = 1,2,\ldots, L_\mathrm{r}.
	\end{align}
	Then, the objective function in (\ref{Opt_3}) can be rewritten as
	\begin{align}\label{Opt_pass}
		\mathbb{E}\left[\left|\mathbf{h}^{\rm{H}} \mathbf{f}_{\mathrm{t}}\right|^{2}\right]
		= \boldsymbol{\psi}^{\rm{H}}{\mathbf{J}_\mathrm{pass}}\boldsymbol{\psi}.
	\end{align}	
	The intermediate variable ${\mathbf{J}_\mathrm{pass}} \in \mathbb{C}^{N_{\mathrm{R}} \times N_{\mathrm{R}}}$ is given by
	\begin{align}\label{J_pass}
		\nonumber {\mathbf{J}_\mathrm{pass}} =
		& \ (4 - 2 Q) \sum\limits_{l_{\mathrm{s}} = 1}^{L_{\mathrm{s}}} \mathbf{B}_{\mathrm{s}, l_{\mathrm{s}}}^{\rm{T}} {{\mathbf{v}}^{*}_{l_\mathrm{s}}} {{\mathbf{v}}^{\rm{T}}_{l_\mathrm{s}}} \mathbf{B}_{\mathrm{s}, l_{\mathrm{s}}}^{*} \\
		\nonumber & + Q \left(\sum\limits_{l_{\mathrm{s}} = 1}^{L_{\mathrm{s}}} \mathbf{B}_{\mathrm{s}, l_{\mathrm{s}}}^{\rm{T}} {{\mathbf{v}}^{*}_{l_\mathrm{s}}} \right) \left( \sum\limits_{l_{\mathrm{s}} = 1}^{L_{\mathrm{s}}} {{\mathbf{v}}^{\rm{T}}_{l_\mathrm{s}}} \mathbf{B}_{\mathrm{s}, l_{\mathrm{s}}}^{*} \right) \\
		\nonumber & + Q \left(\sum\limits_{l_\mathrm{s} = 1}^{L_\mathrm{s}} \mathbf{B}_{\mathrm{s}, l_\mathrm{s}}^{\rm{T}} \boldsymbol{\Upsilon}_{\mathrm{s},l_{\mathrm{s}}}^{\rm{T}} \right) \left(\sum\limits_{l_\mathrm{s} = 1}^{L_\mathrm{s}} \boldsymbol{\Upsilon}_{\mathrm{s},l_{\mathrm{s}}}^{*} \mathbf{B}_{\mathrm{s}, l_\mathrm{s}}^{*} \right) \\
		& + Q \left(\sum\limits_{l_\mathrm{r} = 1}^{L_\mathrm{r}} \mathbf{B}_{\mathrm{r}, l_\mathrm{r}}^{\rm{T}} \boldsymbol{\Upsilon}_{\mathrm{r},l_{\mathrm{r}}}^{\rm{T}} \right) \left(\sum\limits_{l_\mathrm{r} = 1}^{L_\mathrm{r}} \boldsymbol{\Upsilon}_{\mathrm{r},l_{\mathrm{r}}}^{*} \mathbf{B}_{\mathrm{r}, l_\mathrm{r}}^{*} \right),
	\end{align}
	where we define
	\begin{align}
		& \boldsymbol{\Upsilon}_{\mathrm{s}} \! = \! \mathbf{I}_{L_\mathrm{s}} \otimes \mathbf{V}^{\rm{H}} \! = \! \left[\boldsymbol{\Upsilon}_{\mathrm{s}, 1}, \boldsymbol{\Upsilon}_{\mathrm{s}, 2} , \ldots, \boldsymbol{\Upsilon}_{\mathrm{s}, L_\mathrm{s}}\right] \! \in \mathbb{C}^{L_{\mathrm{s}}^{2} \times N_{\mathrm{B}} L_{\mathrm{s}}}, \\
		& \boldsymbol{\Upsilon}_{\mathrm{r}} \! = \! \mathbf{I}_{L_\mathrm{r}} \! \otimes \! \mathbf{V}^{\rm{H}} \! = \! \left[\boldsymbol{\Upsilon}_{\mathrm{r}, 1}, \boldsymbol{\Upsilon}_{\mathrm{r}, 2} , \ldots, \boldsymbol{\Upsilon}_{\mathrm{r}, L_\mathrm{r}}\right] \! \in \! \mathbb{C}^{L_{\mathrm{s}} L_{\mathrm{r}} \times N_{\mathrm{B}} L_{\mathrm{r}}}
	\end{align}
	with $\boldsymbol{\Upsilon}_{\mathrm{s}, l_\mathrm{s}} \in \mathbb{C}^{L_{\mathrm{s}}^{2} \times N_{\mathrm{B}}}$ and $\boldsymbol{\Upsilon}_{\mathrm{r}, l_\mathrm{r}} \in \mathbb{C}^{L_{\mathrm{s}} L_{\mathrm{r}} \times N_{\mathrm{B}}}$.

	In addition, the properties
		\begin{align}
			& \left\|\operatorname{diag} \left(\mathbf{A}^\mathrm{H}_{\mathrm{s}} \mathbf{V}\right)\right\|^{2}
			= \boldsymbol{\psi}^\mathrm{H} \left( \sum\limits_{l_{\mathrm{s}} = 1}^{L_{\mathrm{s}}} \mathbf{B}_{\mathrm{s}, l_{\mathrm{s}}}^\mathrm{T} {{\mathbf{v}}^{*}_{l_\mathrm{s}}} {\mathbf{v}^\mathrm{T}_{l_\mathrm{s}}} \mathbf{B}_{\mathrm{s}, l_{\mathrm{s}}}^{*} \right) \boldsymbol{\psi}, \\
			& \left|\operatorname{tr}(\mathbf{A}^\mathrm{H}_{\mathrm{s}} \mathbf{V})\right|^{2}
			= \boldsymbol{\psi}^\mathrm{H} \left(\sum\limits_{l_{\mathrm{s}} = 1}^{L_{\mathrm{s}}} \mathbf{B}_{\mathrm{s}, l_{\mathrm{s}}}^\mathrm{T} {{\mathbf{v}}^{*}_{l_\mathrm{s}}} \right) \!\! \left( \sum\limits_{l_{\mathrm{s}} = 1}^{L_{\mathrm{s}}} {{\mathbf{v}}^\mathrm{T}_{l_\mathrm{s}}} \mathbf{B}_{\mathrm{s}, l_{\mathrm{s}}}^{*} \right) \boldsymbol{\psi}, \\
			& \left\|\mathbf{A}_{\mathrm{s}}^\mathrm{H} \mathbf{V} \right\|_\mathrm{F}^{2} \!
			= \boldsymbol{\psi}^\mathrm{H} \!\! \left(\sum\limits_{l_\mathrm{s} = 1}^{L_\mathrm{s}} \mathbf{B}_{\mathrm{s}, l_\mathrm{s}}^\mathrm{T} \boldsymbol{\Upsilon}_{\mathrm{s},l_{\mathrm{s}}}^\mathrm{T} \right) \!\! \left(\sum\limits_{l_\mathrm{s} = 1}^{L_\mathrm{s}} \boldsymbol{\Upsilon}_{\mathrm{s},l_{\mathrm{s}}}^{*} \mathbf{B}_{\mathrm{s}, l_\mathrm{s}}^{*} \right) \! \boldsymbol{\psi}, \\
			& \left\|\mathbf{A}_{\mathrm{r}}^\mathrm{H} \mathbf{V} \right\|_\mathrm{F}^{2} 
			\! = \! \boldsymbol{\psi}^\mathrm{H} \left(\sum\limits_{l_\mathrm{r} = 1}^{L_\mathrm{r}} \mathbf{B}_{\mathrm{r}, l_\mathrm{r}}^\mathrm{T} \boldsymbol{\Upsilon}_{\mathrm{r},l_{\mathrm{r}}}^\mathrm{T} \right) \!\! \left(\sum\limits_{l_\mathrm{r} = 1}^{L_\mathrm{r}} \boldsymbol{\Upsilon}_{\mathrm{r},l_{\mathrm{r}}}^{*} \mathbf{B}_{\mathrm{r}, l_\mathrm{r}}^{*} \right) \boldsymbol{\psi}
		\end{align}
	are used here.
	Then, $\mathcal{P}_4$ is equivalently re-expressed as
		\begin{align}\label{Opt_fixedpoint}
			\mathcal{P}_4: \ \max_{\{\boldsymbol{\psi}\}} \boldsymbol{\psi}^{\rm{H}} {\mathbf{J}_\mathrm{pass}} \boldsymbol{\psi}, 
			\ {\text{s.t.}} \ \left|\boldsymbol{\psi}\right| = \mathbf{1}_{N_{\rm{R}} \times 1}.
		\end{align}	
	The solution of $\mathcal{P}_4$ can be obtained via the fixed point iteration \cite{FixedPoint_TSP_2014}.
	In the $(i + 1)$-th iteration, $\boldsymbol{\psi}$ can be updated as 
	\begin{align}\label{Express_fixedpoint}
		\boldsymbol{\psi}^{(i + 1)} = e^{j \arg \left( \mathbf{J}_\mathrm{pass} \boldsymbol{\psi}^{(i)} \right)},
	\end{align}	
	where the local optimality and convergence of (\ref{Express_fixedpoint}) have been proved in \cite{FixedPoint_TSP_2014}.

	\textit{4) Proposed Path Selection Algorithm:}

	After solving the alternating optimization of active and passive beamformers, we perform the path selection.
	In the proposed path selection strategy, to maximize the achievable downlink spectral efficiency $R$, the path with minimal contribution to $R$ is removed sequentially.
	Specifically, according to (\ref{goal_inial}) for $l_{\mathrm{s}} = 1,2,\ldots,L_{\mathrm{s}}$, the component related to the $l_{\mathrm{s}}$-th path is separated from $\mathbb{E}\left[\left|\mathbf{h}^{\rm{H}} \mathbf{f}_{\mathrm{t}}\right|^{2}\right]$ and defined as $\zeta(l_{\mathrm{s}})$, which is given by
	\begin{align}\label{Path Contribution}
		\nonumber \zeta \left( l_{\mathrm{s}} \right)
		& = 4 \mathbf{v}^{\mathrm{H}}_{l_\mathrm{s}} \mathbf{a}_{\mathrm{s}, l_\mathrm{s}} \mathbf{a}_{\mathrm{s}, l_\mathrm{s}}^{\mathrm{H}} \mathbf{v}_{l_\mathrm{s}}
		+ 2 Q \Re \left[ \mathbf{v}_{l_\mathrm{s}}^\mathrm{H} \mathbf{a}_{\mathrm{s}, l_\mathrm{s}} \sum\limits_{i \neq l_\mathrm{s}}^{L_\mathrm{s}} \mathbf{a}_{\mathrm{s}, i}^\mathrm{H} \mathbf{v}_{i} \right] \\
		& + 2 Q \Re \left[ \sum\limits_{i \neq l_\mathrm{s}}^{L_\mathrm{s}} \mathbf{v}_{l_\mathrm{s}}^\mathrm{H} \mathbf{a}_{\mathrm{s}, l_\mathrm{s}} \mathbf{a}_{\mathrm{s}, i}^\mathrm{H} \mathbf{v}_{i} \right]
		+ Q \left\| \mathbf{A}_{\mathrm{r}}^{\mathrm{H}} \mathbf{v}_{l_\mathrm{s}} \right\|^2.
	\end{align}
	The overall path selection algorithm is summarized in Algorithm 1.
	The computational complexity of Algorithm 1 is approximately $\mathcal{O} \left( N^{\mathrm{alt}}_{\mathrm{2}} L^{3} N_{\mathrm{B}} ( N_{\mathrm{fp}} N_{\mathrm{R}} + N_{\mathrm{B}}) \right)$, where $N^{\mathrm{alt}}_{\mathrm{2}}$ denotes the number of alternating iterations for solving $\mathcal{P}_2$, and $N_{\mathrm{fp}}$ denotes the loops required for fixed point iterations.

	\begin{algorithm}[th!]
		\caption{Proposed Path Selection Algorithm.}
		\KwIn{$\mathbf{B}_{l} \in \mathbb{C}^{N_{\mathrm{B}} \times N_{\mathrm{R}}}$ for $l = 1, \ldots, L$ in (\ref{B_defi_2}), $L_{\mathrm{RB}}$, and $L_{\mathrm{RU}}$.}		
		Initialization: $\Lambda_{\mathrm{s}} = \{1,2, \ldots, L\}$; $\boldsymbol{\psi} = \mathbf{1}_{N_{\mathrm{R}} \times 1}$ \\
		\Repeat{$L_{\mathrm{s}}$ \rm{reaches the pre-given number of dominant paths}}
		{
			\Repeat{\rm{the objective value of} $\mathbb{E}\left[\left|\mathbf{h}^{\rm{H}} \mathbf{f}_{\mathrm{t}}\right|^{2}\right]$ \rm{in} $\mathcal{P}_2$ \rm{converges}}
			{
				Compute $\mathbf{A}$, $\mathbf{A}_{\mathrm{s}}$, and $\mathbf{A}_{\mathrm{r}}$ according to $\Lambda_{\mathrm{s}}$, $\boldsymbol{\psi}$, and $\mathbf{B}_{l}$ (see \eqref{A_defi_col}-\eqref{B_defi_2}) \;
				Optimize $\mathbf{V}$ with fixed $\boldsymbol{\psi}$ by solving $\mathcal{P}_3$ \;
				Optimize $\boldsymbol{\psi}$ with fixed $\mathbf{V}$ by solving $\mathcal{P}_4$ (see (\ref{Express_fixedpoint})) \;
			}
			Find the path index denoted as $\hat{l}_{\mathrm{s}}$ with minimal $\zeta(\hat{l}_{\mathrm{s}})$  by (\ref{Path Contribution}) for $\hat{l}_{\mathrm{s}} \in \{1,2, \ldots, L_{\mathrm{s}}\}$ \;
			Update $\{\Lambda_{\mathrm{s}}\}$ by removing its $\hat{l}_{\mathrm{s}}$-th element \;
			$L_{\mathrm{s}} \leftarrow L_{\mathrm{s}} - 1$ \;
		}
		\KwOut{$\Lambda_{\mathrm{s}}$}
	\end{algorithm}

	\subsection{DPGI Estimation and Feedback}\label{S3.2}

	Once the path selection is finished, we perform the acquisition of $\mathbf{g}_{\mathrm{s}} \in \mathbb{C}^{L_{\mathrm{s}} \times 1}$ (downlink DPGI) at BS.
	For $t = 1,2,\ldots,L_{\mathrm{s}}$, pilot symbol denoted as $s(t)$ is sent at BS within $L_{\mathrm{s}}$ time slots.
	Then, based on (\ref{h_cascaded}), the signal received at UE in the $t$-th time slot is given by
	\begin{align}\label{y_ce_inial}
		\nonumber y(t) = & \ \sqrt{\frac{P_{\mathrm{e}} N_{\mathrm{B}} N_{\mathrm{R}}^{2}}{L}} \boldsymbol{\beta}^{\rm{T}} \mathbf{A}_{\mathrm{RU}}^{\rm{H}} \operatorname{diag}(\boldsymbol{\psi}(t)) \mathbf{A}_{\mathrm{RB}} \\
		& \times \operatorname{diag}(\boldsymbol{\alpha}) \mathbf{A}_{\mathrm{B}}^{\rm{H}} \mathbf{f}_{\mathrm{e}}(t) s(t) + n(t),
	\end{align}
	where $P_{\mathrm{e}}$ is the transmitting power for DPGI estimation, $n(t) \sim \mathcal{C N}\left(0, \sigma_{n}^{2}\right)$ is the complex Gaussian noise with noise power $\sigma_{n}^{2}$.
	In addition, $\mathbf{f}_{\mathrm{e}} (t) \in \mathbb{C}^{N_\mathrm{B} \times 1}$ satisfying $\left\| \mathbf{f}_{\mathrm{e}}(t) \right\|^{2} = 1$ represents the active beamforming vector for DPGI estimation (different from the $\mathbf{f}_{\mathrm{t}}$ defined in (\ref{f_Vg}) for data transmission), and $\boldsymbol{\psi}(t)$ denotes the passive beamforming vector in the $t$-th time slot for DPGI estimation satisfying $\left|\boldsymbol{\psi}(t)\right| = \mathbf{1}_{N_{\rm{R}} \times 1}$.
	The performance of DPGI estimation can be improved by jointly designing the active and passive beamformers.
	By assuming $s(t) = 1$, for $t = 1,2,\ldots,L_{\mathrm{s}}$, the received signal $y(t)$ in (\ref{y_ce_inial}) is re-expressed as
	\begin{align}
		y(t) & = \sqrt{\frac{P_{\mathrm{e}} N_{\mathrm{B}} N_{\mathrm{R}}^{2}}{L}} \boldsymbol{\psi}^{\rm{T}}(t) \tilde{\mathbf{A}}_{\mathrm{R}} \operatorname{diag}(\mathbf{g}) \tilde{\mathbf{A}}_{\mathrm{B}}^{\rm{H}} \mathbf{f}_{\mathrm{e}}(t) + n(t) \label{y_ce_1} \\
		& = \sqrt{ \! \frac{P_{\mathrm{e}} N_{\mathrm{B}} N_{\mathrm{R}}^{2}}{L}} \! \left( \! \left( \! \mathbf{f}_{\mathrm{e}}^{\rm{T}}(t) \tilde{\mathbf{A}}_{\mathrm{B}}^{*} \! \right) \! \odot \! \left( \! \boldsymbol{\psi}^{\rm{T}} \! (t) \tilde{\mathbf{A}}_{\mathrm{R}} \! \right) \! \right) \mathbf{g} \! + \! n(t) \label{y_ce_2} \\
		& = \boldsymbol{\kappa}^{\rm{T}}(t) \mathbf{g} + n(t) \label{y_ce_3} \\
		& = \underbrace{\kappa_{\lambda_{\mathrm{s},t}}(t) g_{\lambda_{\mathrm{s},t}}}_{\mathrm{To \ be \ maximized}} + \underbrace{\left( \sum\limits_{l \neq \lambda_{\mathrm{s},t}}^{L} \kappa_{l}(t) g_l + n(t) \right)}_{\mathrm{Equivalent \ noise}}, \label{y_ce_4}
	\end{align}
	where (\ref{y_ce_1}) is obtained by defining
	\begin{align}\label{A_e}
		& \tilde{\mathbf{A}}_{\mathrm{B}} = [\tilde{\mathbf{a}}_{\mathrm{B}, 1}, \ldots , \tilde{\mathbf{a}}_{\mathrm{B}, L}] \in \mathbb{C}^{N_{\mathrm{B}} \times L}, \tilde{\mathbf{a}}_{{\mathrm{B}}, l} = \mathbf{a}_{\mathrm{B}, p}, \\
		& \tilde{\mathbf{A}}_{\mathrm{R}} \! = \! [\tilde{\mathbf{a}}_{\mathrm{R}, 1}, \! \ldots \! , \! \tilde{\mathbf{a}}_{\mathrm{R}, L}] \!\! \in \! \mathbb{C}^{N_{\mathrm{R}} \! \times \! L} \! , \tilde{\mathbf{a}}_{\mathrm{R}, l} \! = \! \operatorname{diag} \! \left( \! \mathbf{a}_{\mathrm{RU}, q}^{*} \right) \! \mathbf{a}_{\mathrm{RB}, p}
	\end{align}
	for the path index $l = (q - 1) L_{\mathrm{RB}} + p$, $p = 1, 2, \ldots, L_{\mathrm{RB}}$ and $q = 1, 2, \ldots, L_{\mathrm{RU}}$.
	Besides, (\ref{y_ce_2}) follows from the property of Hadamard product, and $\boldsymbol{\kappa}(t) \in \mathbb{C}^{L \times 1}$ is defined as
	\begin{align}\label{kappa}
		\nonumber \boldsymbol{\kappa}(t) = & \ [\kappa_{1}(t), \kappa_{2}(t), \ldots, \kappa_{L}(t)]^{\rm{T}} \\
		= & \sqrt{\frac{P_{\mathrm{e}} N_{\mathrm{B}} N_{\mathrm{R}}^{2}}{L}} \left( \left( \mathbf{f}^{\rm{T}}_{\mathrm{e}}(t) \tilde{\mathbf{A}}_{\mathrm{B}}^{*} \right) \odot \left( \boldsymbol{\psi}^{\rm{T}}(t) \tilde{\mathbf{A}}_{\mathrm{R}} \right) \right)^{\rm{T}}
	\end{align}
	in (\ref{y_ce_3}) to simplify the expression.
	The strategy of DPGI estimation can be explained using (\ref{y_ce_4}).
	Specifically, for the selected $L_{\mathrm{s}}$ dominant paths, the corresponding active and passive beamforming vectors are jointly optimized in each time slot $t = 1,2,\ldots, L_{\mathrm{s}}$ to maximize $\kappa_{\lambda_{\mathrm{s},t}}(t)$, where $\lambda_{\mathrm{s},t}$ denotes the $t$-th element in set $\Lambda_{\mathrm{s}} = \{\lambda_{\mathrm{s},1}, \lambda_{\mathrm{s},2}, \ldots, \lambda_{\mathrm{s},L_{\mathrm{s}}}\}$.
	Then, the component corresponding to the $t$-th dominant path in (\ref{y_ce_4}) denoted as $\kappa_{\lambda_{\mathrm{s},t}}(t) g_{\lambda_{\mathrm{s},t}}$ is maximized, and the rest components are considered as the equivalent noise.

	Based on (\ref{y_ce_2}), the maximization of $\kappa_{\lambda_{\mathrm{s},t}}(t)$ is equivalent to the maximization of the $\lambda_{\mathrm{s},t}$-th element in row vector $\mathbf{f}_{\mathrm{e}}^{\rm{T}}(t) \tilde{\mathbf{A}}_{\mathrm{B}}^{*} \in \mathbb{C}^{1 \times L}$ and the maximization of the $\lambda_{\mathrm{s},t}$-th element in row vector $\boldsymbol{\psi}^{\rm{T}}(t) \tilde{\mathbf{A}}_{\mathrm{R}} \in \mathbb{C}^{1 \times L}$.
	Therefore, the optimal solutions of jointly designed active beamformer $\mathbf{f}_{\mathrm{e}} (t)$ and passive beamformer $\boldsymbol{\psi} (t)$ are given, respectively, by
	\begin{align}
		& \mathbf{f}_{\mathrm{e}}(t) = \frac{\tilde{\mathbf{a}}_{\mathrm{B},\lambda_{\mathrm{s},t}}} {\left\| \tilde{\mathbf{a}}_{\mathrm{B},\lambda_{\mathrm{s},t}} \right\|} = \tilde{\mathbf{a}}_{\mathrm{B},\lambda_{\mathrm{s},t}},  \label{act_ce} \\
		& \boldsymbol{\psi}(t) = \frac{\tilde{\mathbf{a}}^{*}_{\mathrm{R},\lambda_{\mathrm{s},t}}} {| \tilde{\mathbf{a}}^{*}_{\mathrm{R},\lambda_{\mathrm{s},t}} |} = N_{\mathrm{R}} \tilde{\mathbf{a}}^{*}_{\mathrm{R},\lambda_{\mathrm{s},t}}, \label{pass_ce}
	\end{align} 
	where we have $\left\| \tilde{\mathbf{a}}_{\rm{B},\lambda_{\mathrm{s},t}} \right\| = 1$ and
	$| \tilde{\mathbf{a}}^{*}_{\mathrm{R},\lambda_{\mathrm{s},t}} | = \frac{1}{N_{\mathrm{R}}} \mathbf{1}_{N_{\mathrm{R}} \times 1}$ according to their definitions.
	The power constraint and the constant modulus constraint are satisfied in (\ref{act_ce}) and (\ref{pass_ce}), respectively.
	By substituting (\ref{act_ce}) and (\ref{pass_ce}) into (\ref{kappa}), we obtain
	\begin{align}\label{g_coefficient}
		\kappa_{\lambda_{\mathrm{s},t}} \! (t) \! = \!\! \sqrt{ \! \frac{P_{\mathrm{e}} N_{\mathrm{B}} N_{\mathrm{R}}^{2}}{L}} \mathbf{f}_{\mathrm{e}}^{\rm{T}} \! (t) \tilde{\mathbf{a}}^{*}_{\mathrm{B},\lambda_{\mathrm{s},t}} \! \boldsymbol{\psi}^{\rm{T}} \! (t) \tilde{\mathbf{a}}_{\mathrm{R},\lambda_{\mathrm{s},t}} \!\! = \!\! \sqrt{ \! \frac{P_{\mathrm{e}} N_{\mathrm{B}} N_{\mathrm{R}}^{2}}{L}}. \!
	\end{align}
	We further have $\mathbf{y} = [y(1), y(2), \ldots, y(L_{\mathrm{s}})]^{\rm{T}} \in \mathbb{C}^{L_{\mathrm{s}} \times 1}$ by concatenating $L_{\mathrm{s}}$ successive receiving signals.
	According to (\ref{y_ce_4}) and (\ref{g_coefficient}), the estimate of downlink DPGI at the UE is given by
	\begin{align}
		\hat{\mathbf{g}}_{\mathrm{s}} = \sqrt{\frac{L} {P_{\mathrm{e}} N_{\mathrm{B}} N_{\mathrm{R}}^{2}}} {\mathbf{y}} \in \mathbb{C}^{L_{\mathrm{s}} \times 1}.
	\end{align}
	Then, the RVQ codebook is adopted i.e., $\mathbf{C}_{\mathrm{RVQ}} =\left[\mathbf{c}_{1}, \mathbf{c}_{2}, \ldots, \mathbf{c}_{2^{B}}\right] \in \mathbb{C}^{L_{\mathrm{s}} \times 2^{B}}$ with $B$ bits for the uplink feedback of $\hat{\mathbf{g}}_{\mathrm{s}} \in \mathbb{C}^{L_{\mathrm{s}} \times 1}$.
	By normalizing the estimated DPGI as $\bar{\hat{\mathbf{g}}}_{\mathrm{s}} = \frac{\hat{\mathbf{g}}_{\mathrm{s}}}{\left\|\hat{\mathbf{g}}_{\mathrm{s}}\right\|} \in \mathbb{C}^{L_{\mathrm{s}} \times 1}$ and choosing a corresponding feedback codeword $\hat{i}$ according to (\ref{RVQ_defi}), the quantized downlink DPGI acquired at BS is expressed as
	\begin{align}\label{g_s_qua_est}
		\tilde{\mathbf{g}}_{\mathrm{s}} = \left\|\hat{\mathbf{g}}_{\mathrm{s}}\right\| \mathbf{c}_{\hat{i}}.
	\end{align}

	\subsection{Beamformers Update}\label{S3.3}

	Recall that the active beamforming matrix $\mathbf{V}$ and passive beamforming vector $\boldsymbol{\psi}$ are designed in the path selection step. 
	Note that the expression of $\mathbb{E}\left[\left|\mathbf{h}^{\rm{H}} \mathbf{f}_{\mathrm{t}}\right|^{2}\right]$ in (\ref{goal_inial}) is derived based on the statistical information of PGI, including $\mathbf{g}_{\mathrm{s}} \in \mathbb{C}^{L_{\mathrm{s}} \times 1}$.
	In this subsection, to further improve the spectral efficiency, we update the passive beamformer $\boldsymbol{\psi}$ and the active beamformer $\mathbf{f}_{\mathrm{t}}$ by exploiting the quantized DPGI $\tilde{\mathbf{g}}_{\mathrm{s}}$.$\footnote{Since the quantized downlink DPGI can be used for active and passive beamformers update, here we can adopt the direct optimizition of $\mathbf{f}_{\mathrm{t}}$ instead of the indirect optimization of $\mathbf{f}_{\mathrm{t}}$ by solving $\mathbf{V}$ according to $\mathbf{f}_{\mathrm{t}} = \mathbf{V} \mathbf{g}^{*}_{\mathrm{s}}$ as defined in (\ref{f_Vg}).}$

	\textit{1) Problem Formulation:}

	After performing the proposed path selection algorithm together with DPGI estimation and feedback scheme, the index set of selected dominant paths $\{\Lambda_{\mathrm{s}}\}$ and the quantized DPGI $\tilde{\mathbf{g}}_{\mathrm{s}} \in \mathbb{C}^{L_{\mathrm{s}} \times 1}$ are acquired at BS.
	Then, we further update the optimization problem based on $\mathcal{P}_1$ as
	\begin{align}\label{Opt_5}
		\mathcal{P}_5: \ \max_{\{\mathbf{f}_{\mathrm{t}}, \boldsymbol{\psi}\}} \mathbb{E}\left[\left|\mathbf{h}^{\rm{H}} \mathbf{f}_{\mathrm{t}}\right|^{2}\right],
		\ {\text{s.t.}} \ \left\|\mathbf{f}_{\mathrm{t}}\right\|^{2} = 1,
		\left|\boldsymbol{\psi}\right| = \mathbf{1}_{N_{\rm{R}} \times 1}.
	\end{align}
	Compared to the statistical information of $\mathbf{g}_{\mathrm{s}}$ used in $\mathcal{P}_2$, the quantized DPGI $\tilde{\mathbf{g}}_{\mathrm{s}}$ obtained through uplink feedback can be utilized in $\mathcal{P}_5$, resulting in a different formulation of the optimization problem. Therefore, we additionally perform an alternating optimization between the updates of $\mathbf{f}_{\mathrm{t}}$ and $\boldsymbol{\psi}$ as below.

	\textit{2) Active Beamformer Update:}

	On the basis of $\mathcal{P}_5$, the optimization of $\mathbf{f}_{\mathrm{t}}$ with fixed $\boldsymbol{\psi}$ is formulated as
	\begin{align}\label{Opt_6}
		\mathcal{P}_6: \ \max_{\{\mathbf{f}_{\mathrm{t}}\}} \mathbb{E}\left[\left|\mathbf{h}^{\rm{H}} \mathbf{f}_{\mathrm{t}}\right|^{2}\right], 
		\ {\text{s.t.}} \ \left\|\mathbf{f}_{\mathrm{t}}\right\|^{2} = 1.
	\end{align}	

	\noindent \textbf{Lemma 2:}	
	We can further re-express $\mathbb{E}\left[\left|\mathbf{h}^{\rm{H}} \mathbf{f}_{\mathrm{t}}\right|^{2}\right]$ in (\ref{Opt_6}) as
	\begin{align}\label{goal_update}
		\mathbb{E}\left[\left|\mathbf{h}^{\rm{H}} \mathbf{f}_{\mathrm{t}}\right|^{2}\right]
		= \left\|\mathbf{A}_{\mathrm{r}}^{\rm{H}} \mathbf{f}_{\mathrm{t}} \right\|^{2} + \left|\mathbf{g}_{\mathrm{s}}^{\rm{T}} \mathbf{A}_{\mathrm{s}}^{\rm{H}} \mathbf{f}_{\mathrm{t}} \right|^{2},
	\end{align}	
	where $\mathbf{g}_{\mathrm{s}}$ is kept because BS has obtained its quantized estimate $\tilde{\mathbf{g}}_{\mathrm{s}}$, and $\mathbf{g}_{\mathrm{r}}$ is removed in the final expression of expectation by using the statistical information of $\mathbf{g}_{\mathrm{r}}$, which makes (\ref{goal_update}) different from (\ref{goal_inial}) obtained in Lemma 1.
	\begin{IEEEproof}[Proof]
		See Appendix B.
	\end{IEEEproof}

	The objective function in (\ref{goal_update}) is further expressed as
	\begin{align}
		\mathbb{E}\left[\left|\mathbf{h}^{\rm{H}} \mathbf{f}_{\mathrm{t}}\right|^{2}\right]
		= \mathbf{f}_{\mathrm{t}}^{\rm{H}} \tilde{\mathbf{J}}_\mathrm{act} \mathbf{f}_{\mathrm{t}},
	\end{align}	
	where $\tilde{\mathbf{J}}_\mathrm{act}$ is expressed as
	\begin{align}
		\tilde{\mathbf{J}}_\mathrm{act} = \mathbf{A}_{\mathrm{r}} \mathbf{A}_{\mathrm{r}}^{\rm{H}} + \mathbf{A}_{\mathrm{s}} \mathbf{g}_{\mathrm{s}}^{*} \mathbf{g}_{\mathrm{s}}^{\rm{T}} \mathbf{A}_{\mathrm{s}}^{\rm{H}} \in \mathbb{C}^{N_{\mathrm{B}} \times N_{\mathrm{B}}}.
	\end{align}	
	Then, we obtain the optimal solution $\mathbf{f}_{\mathrm{t}}^{\star}$ of $\mathcal{P}_6$ as
	\begin{align}
		\mathbf{f}_{\mathrm{t}}^{\star} = \frac{ \tilde{\mathbf{u}}_{\mathrm{act}, max} }{\left\| \tilde{\mathbf{u}}_{\mathrm{act}, max} \right\|} \in \mathbb{C}^{N_{\mathrm{B}} \times 1},
	\end{align}
	where $\tilde{\mathbf{u}}_{\mathrm{act}, max}$ denotes the eigenvector corresponding to the largest eigenvalue of ${\tilde{\mathbf{J}}_\mathrm{act}}$.

	\textit{3) Passive Beamformer Update:}

	On the basis of $\mathcal{P}_5$, the optimization of $\boldsymbol{\psi}$ with fixed $\mathbf{f}_{\mathrm{t}}$ can be formulated as
	\begin{align}\label{Opt_7}
		\mathcal{P}_7: \ \max_{\{\boldsymbol{\psi}\}} \boldsymbol{\psi}^{\rm{H}} {\tilde{\mathbf{J}}_\mathrm{pass}} \boldsymbol{\psi}, 
		\ {\text{s.t.}} \ \left|\boldsymbol{\psi}\right| = \mathbf{1}_{N_{\rm{R}} \times 1}, 
	\end{align}
	where we have
	\begin{align}\label{Tilde_J_pass}
		\nonumber \tilde{\mathbf{J}}_\mathrm{pass} & = \left(\sum\limits_{l_\mathrm{r} = 1}^{L_\mathrm{r}} \mathbf{B}_{\mathrm{r}, l_\mathrm{r}}^{\rm{T}} \boldsymbol{\Gamma}_{\mathrm{r}, l_{\mathrm{r}}}^{\rm{T}} \right) \left( \sum\limits_{l_\mathrm{r} = 1}^{L_\mathrm{r}} \boldsymbol{\Gamma}_{\mathrm{r}, l_{\mathrm{r}}}^{*} \mathbf{B}_{\mathrm{r}, l_\mathrm{r}}^{*} \right) \\
		& + \! \left(\sum\limits_{l_\mathrm{s} = 1}^{L_\mathrm{s}} \mathbf{B}_{\mathrm{s}, l_\mathrm{s}}^{\rm{T}} \boldsymbol{\gamma}^{*}_{\mathrm{s}, l_\mathrm{s}} \right) \!\! \left( \sum\limits_{l_\mathrm{s} = 1}^{L_\mathrm{s}} \boldsymbol{\gamma}^{\rm{T}}_{\mathrm{s}, l_\mathrm{s}} \mathbf{B}_{\mathrm{s}, l_\mathrm{s}}^{*} \right) \! \in \! \mathbb{C}^{N_{\mathrm{R}} \times N_{\mathrm{R}}}.
	\end{align}
	The intermediate variables $\boldsymbol{\Gamma}_{\mathrm{r}}$ and $\boldsymbol{\gamma}_{\mathrm{s}}$ are defined as 
	\begin{align}
		& \boldsymbol{\Gamma}_{\mathrm{r}} = \mathbf{I}_{L_{\mathrm{r}}} \otimes \mathbf{f}_{\mathrm{t}}^{\rm{H}} = \left[ \boldsymbol{\Gamma}_{\mathrm{r},1}, \boldsymbol{\Gamma}_{\mathrm{r},2} , \ldots, \boldsymbol{\Gamma}_{\mathrm{r}, L_\mathrm{r}} \right] \in \mathbb{C}^{ L_{\mathrm{r}} \times N_{\mathrm{B}} L_{\mathrm{r}}}, \\
		& \boldsymbol{\gamma}_{\mathrm{s}} = \mathbf{g}_{\mathrm{s}} \otimes \mathbf{f}_{\mathrm{t}} = \left[\boldsymbol{\gamma}^{\rm{T}}_{\mathrm{s}, 1}, \boldsymbol{\gamma}^{\rm{T}}_{\mathrm{s}, 2} , \ldots, \boldsymbol{\gamma}^{\rm{T}}_{\mathrm{s}, L_\mathrm{s}} \right]^{\rm{T}} \in \mathbb{C}^{ N_{\mathrm{B}} L_{\mathrm{s}} \times 1},
	\end{align} 
	where we have $\boldsymbol{\Gamma}_{\mathrm{r}, l_\mathrm{r}} \in \mathbb{C}^{ L_{\mathrm{r}} \times N_{\mathrm{B}}}$ and $\boldsymbol{\gamma}_{\mathrm{s}, l_\mathrm{s}} \in \mathbb{C}^{ N_{\mathrm{B}} \times 1}$ for $l_{\mathrm{r}} = 1,2,\ldots, L_\mathrm{r}$ and $l_{\mathrm{s}} = 1,2,\ldots, L_\mathrm{s}$, respectively.
	Besides, the properties
	\begin{align}
		\nonumber & \left\| \mathbf{A}_{\mathrm{r}}^\mathrm{H} \mathbf{f}_{\mathrm{t}} \right\|^{2} \!\!
		= \left\| \boldsymbol{\Gamma}_{\mathrm{r}} \operatorname{vec}(\mathbf{A}_{\mathrm{r}}) \right\|^{2} \!\!
		= \left\| \sum\limits_{l_\mathrm{r} = 1}^{L_\mathrm{r}} \boldsymbol{\Gamma}_{\mathrm{r}, l_{\mathrm{r}}}  \mathbf{B}_{\mathrm{r}, l_\mathrm{r}} \boldsymbol{\psi}^{*} \right\|^{2} \\
		& \qquad \quad \ \  = \boldsymbol{\psi}^\mathrm{H} \!\! \left(\sum\limits_{l_\mathrm{r} = 1}^{L_\mathrm{r}} \mathbf{B}_{\mathrm{r}, l_\mathrm{r}}^\mathrm{T} \boldsymbol{\Gamma}_{\mathrm{r}, l_{\mathrm{r}}}^\mathrm{T} \right) \!\! \left( \sum\limits_{l_\mathrm{r} = 1}^{L_\mathrm{r}} \boldsymbol{\Gamma}_{\mathrm{r}, l_{\mathrm{r}}}^{*} \mathbf{B}_{\mathrm{r}, l_\mathrm{r}}^{*} \right) \!\! \boldsymbol{\psi}, \\
		\nonumber & \left|\mathbf{g}_{\mathrm{s}}^\mathrm{T} \mathbf{A}_{\mathrm{s}}^\mathrm{H} \mathbf{f}_{\mathrm{t}}\right|^{2} \!\!
		= \left| \boldsymbol{\gamma}^\mathrm{H}_{\mathrm{s}} \operatorname{vec}(\mathbf{A}_{\mathrm{s}}) \right|^{2} \!\!
		= \left| \sum\limits_{l_\mathrm{s} = 1}^{L_\mathrm{s}} \boldsymbol{\gamma}^\mathrm{H}_{\mathrm{s}, l_\mathrm{s}} \mathbf{B}_{\mathrm{s}, l_{\mathrm{s}}} \boldsymbol{\psi}^{*}\right|^{2} \\
		& \qquad \quad \ \ \ \ = \boldsymbol{\psi}^\mathrm{H} \!\! \left(\sum\limits_{l_\mathrm{s} = 1}^{L_\mathrm{s}} \mathbf{B}_{\mathrm{s}, l_\mathrm{s}}^\mathrm{T} \boldsymbol{\gamma}^{*}_{\mathrm{s}, l_\mathrm{s}} \right) \!\! \left( \sum\limits_{l_\mathrm{s} = 1}^{L_\mathrm{s}} \boldsymbol{\gamma}^\mathrm{T}_{\mathrm{s}, l_\mathrm{s}} \mathbf{B}_{\mathrm{s}, l_\mathrm{s}}^{*} \right) \!\! \boldsymbol{\psi}
	\end{align}	
	are used here.
	By replacing the $\mathbf{J}_\mathrm{pass}$ in (\ref{Express_fixedpoint}) with the $\tilde{\mathbf{J}}_\mathrm{pass}$ in (\ref{Tilde_J_pass}), the fixed point iterations can be employed to solve $\mathcal{P}_7$, which ensures the guaranteed convergence for optimizing the passive beamformer $\boldsymbol{\psi}$ \cite{FixedPoint_TSP_2014}.
	
	\textit{4) Proposed Beamformers Update Algorithm:}

	For the proposed beamformers update algorithm (summarized in Algorithm 2), the overall computational complexity is approximately $\mathcal{O} \left( N^{\mathrm{alt}}_{\mathrm{5}} L N_{\mathrm{B}} ( N_{\mathrm{fp}} N_{\mathrm{R}} + N_{\mathrm{B}}) \right)$, where $N^{\mathrm{alt}}_{\mathrm{5}}$ denotes the number of alternating iterations for solving $\mathcal{P}_5$, and $N_{\mathrm{fp}}$ denotes the loops required for fixed point iterations.

	\begin{algorithm}[th!]
		\caption{Proposed Beamformers Update Algorithm.}
		\KwIn{$\mathbf{B}_{l} \in \mathbb{C}^{N_{\mathrm{B}} \times N_{\mathrm{R}}}$ for $l = 1, \ldots, L$ in (\ref{B_defi_2}), $\{\Lambda_{\mathrm{s}}\}$, and $\tilde{\mathbf{g}}_{\mathrm{s}}$.}		
		Initialization: $\boldsymbol{\psi} = \mathbf{1}_{N_{\mathrm{R}} \times 1}$ \\
		\Repeat{\rm{the objective value of} $\mathbb{E}\left[\left|\mathbf{h}^{\rm{H}} \mathbf{f}_{\mathrm{t}}\right|^{2}\right]$ \rm{in} $\mathcal{P}_5$ \rm{converges}}
		{
			Compute $\mathbf{A}$, $\mathbf{A}_{\mathrm{s}}$, and $\mathbf{A}_{\mathrm{r}}$ according to $\Lambda_{\mathrm{s}}$, $\boldsymbol{\psi}$, and $\mathbf{B}_{l}$ for $l = 1, \ldots, L$ (see \eqref{A_defi_col}-\eqref{B_defi_2}) \;
			Update $\mathbf{f}_{\mathrm{t}}$ with fixed $\boldsymbol{\psi}$ by solving $\mathcal{P}_6$ \;
			Update $\boldsymbol{\psi}$ with fixed $\mathbf{f}_{\mathrm{t}}$ by solving $\mathcal{P}_7$ \;
		}
		\KwOut{The updated active beamformer $\mathbf{f}_{\mathrm{t}}$ and passive beamformer $\boldsymbol{\psi}$}
	\end{algorithm}

    \section{Simulation Results}\label{S4}

	In this section, we evaluate the performance of proposed algorithms.
	In our simulations, we consider the FDD-based RIS-assisted mmWave wireless communication system, where the carrier frequencies of the downlink and uplink are set to $f_{\mathrm{DL}} = 27.875$ GHz and $f_{\mathrm{UL}} = 28.125$ GHz, respectively, and each link is allocated with 250 MHz bandwidth \cite{FDD_mmWaveFreqBand_TVT_2018}.
	The noise power is calculated by $\sigma^{2}_{\mathrm{n}} = -174 + 10 \log_{10} B \approx -90$ dBm.
	The BS is equipped with $N_{\mathrm{B}} = N_{\mathrm{B,v}} \times N_{\mathrm{B,h}} = 4 \times 4$ antennas, and UE is equipped with single antenna.
	Besides, an RIS is equipped with $N_{\mathrm{R}} = N_{\mathrm{R,v}} \times N_{\mathrm{R,h}} = 16 \times 16$ reflecting elements.
	We adopt the channel model and signal model in (\ref{h_cascaded}) and (\ref{y_dt}) with $L_{\mathrm{RB}} = 2$, $L_{\mathrm{RU}} = 3$, where the AoAs/AoDs are assumed to be uniformly distributed in $(0, \pi]$.
	In this paper, we define two types of signal-to-noise ratio (SNR) \cite{LowPS_GXC_TWC_2022}.
	The pilot-to-noise ratio (PNR) is defined as $10 \log _{10} \left(P_{\mathrm{e}} / \sigma_{\mathrm{n}}^2\right)$, where $P_{\mathrm{e}}$ denotes the transmitting power for DPGI estimation in (\ref{y_ce_inial}).
	Similarly, the data-to-noise ratio (DNR) is defined as $10 \log _{10} \left(P_{\mathrm{t}} / \sigma_{\mathrm{n}}^2\right)$, where $P_{\mathrm{t}}$ represents the transmitting power for data transmission in (\ref{y_dt}).
	All results are obtained over 1,000 randomly generated realizations to avoid overfitting to special scenarios.
	
	\subsection{Performance of DPGI Estimation and Feedback}\label{S4.1}

	In this subsection, the proposed DPGI estimation and feedback scheme are compared with several estimators based on existing schemes.
    To the best of our knowledge, there has been no attempt to estimate the DPGI ${\mathbf{g}}_{\mathrm{s}} \in \mathbb{C}^{L_{\mathrm{s}} \times 1}$ (rather than the whole PGI ${\mathbf{g}} \in \mathbb{C}^{L \times 1}$) based on the known PAI in RIS-assisted systems.
	For this reason, to form meaningful comparisons, we not only consider the proposed scenario of $L_{\mathrm{s}} < L$, but also the scenario of $L_{\mathrm{s}} = L$, where all the cascaded paths are used for the dominant paths.
	The LS-based estimator in \cite{CE_CS_WangPeilian_Lett_2020} and the MMSE-based estimator in \cite{Book_Statistical_Signal_Processing} (hereinafter referred to as the LS estimator and MMSE estimator) are used for the estimation of PGI ${\mathbf{g}} \in \mathbb{C}^{L \times 1}$ to form effective comparisons with the proposed DPGI estimation and feedback scheme when $L_{\mathrm{s}} = L = 6$.
	
	Here we provide the processes of LS estimator and MMSE estimator as follows.
	Based on the signal model in (\ref{y_ce_1}), downlink pilot signals are sent from BS within $T$ successive time slots, where $s(t) = 1$ for $t = 1,2,\ldots,T$. The signal received at UE is given by $y(t) = \mathbf{d}^{\rm{T}}(t) \mathbf{g} + n(t)$, where we define
	\begin{align}\label{d_ce}
		\nonumber \mathbf{d}(t) & = [d_{1}(t), d_{2}(t), \ldots, d_{L}(t)]^{\rm{T}} \\
		& = \! \sqrt{ \! \frac{P_{\mathrm{e}} N_{\mathrm{B}} N_{\mathrm{R}}^{2}}{L}} \! \left( \! \left( \mathbf{f}^{\rm{T}}_{\mathrm{e}} \! (t) \tilde{\mathbf{A}}_{\mathrm{B}}^{*} \! \right) \! \odot \! \left( \! \boldsymbol{\psi}^{\rm{T}} \! (t) \tilde{\mathbf{A}}_{\mathrm{R}} \! \right) \! \right)^{\! \rm{T}} \!\!\! \in \! \mathbb{C}^{L \times 1}. \!
	\end{align}
	Different from the $\boldsymbol{\kappa}(t)$ defined in (\ref{kappa}), here we additionally define $\mathbf{d}(t)$ because the active beamforming vector $\mathbf{f}_{\mathrm{e}}(t)$ and the passive beamforming vector $\boldsymbol{\psi}(t)$ used in LS estimator and MMSE estimator are randomly generated.
	By concatenating the signals within $T$ time slots, we have $\mathbf{y} = \mathbf{D}^{\rm{T}} \mathbf{g} + \mathbf{n}$, where $\mathbf{y} = [y(1), y(2), \ldots, y(T)]^{\rm{T}} \in \mathbb{C}^{T \times 1}$, $\mathbf{D} = [\mathbf{d}(1), \mathbf{d}(2), \ldots, \mathbf{d}(T)] \in \mathbb{C}^{L \times T}$, and $\mathbf{n} = [n(1), n(2), \ldots, n(T)]^{\rm{T}} \in \mathbb{C}^{T \times 1}$.
	We note that BS estimates the PAI through the uplink pilot signals, as discussed in Subsection \ref{S2.2}. 
	However, the PAI is not available at UE, and therefore, the matrix $\mathbf{D}$ containing the PAI cannot be obtained at UE.
	For this reason, UE cannot estimate the PGI directly, and hence the received signals $\mathbf{y}$ needs to be fed back via uplink and the PGI estimation is performed at BS, where $T$ (the number of pilot signals) must be greater than or equal to $L$ (the length of PGI) to avoid under-determination estimations \cite{Book_Statistical_Signal_Processing}.
	Then, the RVQ codebook defined in (\ref{RVQ_defi}) is used for the uplink feedback of $\mathbf{y}$, and LS estimator and MMSE estimator are performed at BS by exploiting the quantized vector of received signals (denoted as $\tilde{\mathbf{y}}$), which are given by $\hat{\mathbf{g}}_{\mathrm{LS}} = (\mathbf{D}^{\rm{*}} \mathbf{D}^{\rm{T}})^{-1} \mathbf{D}^{\rm{*}} \tilde{\mathbf{y}} \in \mathbb{C}^{L \times 1}$ and $\hat{\mathbf{g}}_{\mathrm{MMSE}} = \mathbf{D}^{\rm{*}} (\mathbf{D}^{\rm{T}} \mathbf{D}^{\rm{*}} + \sigma^{2}_{\mathrm{n}} \mathbf{I}_{T})^{-1} \tilde{\mathbf{y}} \in \mathbb{C}^{L \times 1}$ \cite{CE_CS_WangPeilian_Lett_2020,Book_Statistical_Signal_Processing,Xing_Matrix_Monotonic_MIMO}.

	\begin{figure}[t]
		\center{\includegraphics[width=0.5\textwidth]{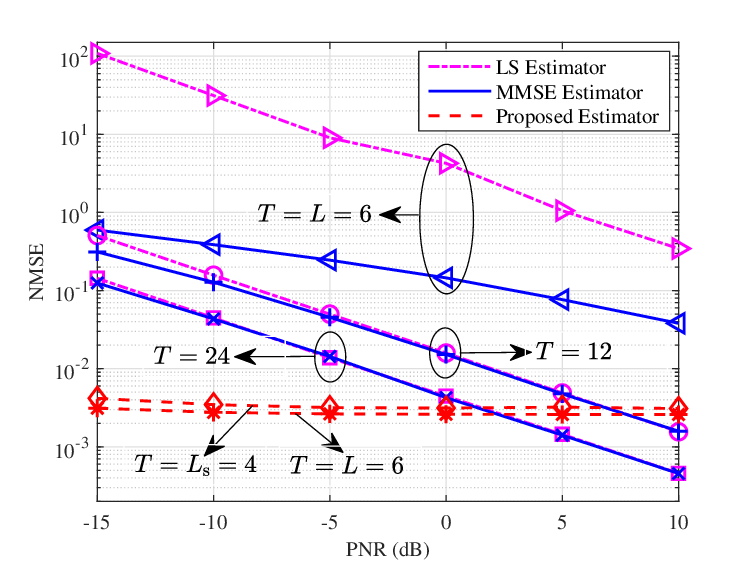}}
		\caption{NMSE performance against PNR when $N_{\mathrm{B}} = N_{\mathrm{B,v}} \times N_{\mathrm{B,h}} = 4 \times 4$, $N_{\mathrm{R}} = N_{\mathrm{R,v}} \times N_{\mathrm{R,h}} = 16 \times 16$, $L = L_{\mathrm{RB}} \times L_{\mathrm{RU}} = 2 \times 3$.}
		\label{fig2}
	\end{figure}

	To evaluate the accuracy of DPGI/PGI estimation, the normalized mean square errors (NMSEs) against PNR are shown in Fig.~\ref{fig2}.
	For the scenario of $L_{\mathrm{s}} < L$, NMSE is expressed as $\mathbb{E} \left[ \| \tilde{\mathbf{g}}_{\mathrm{s}} - \mathbf{g}_{\mathrm{s}} \|^2 / \|\mathbf{g}_{\mathrm{s}}\|^2 \right]$, where $\tilde{\mathbf{g}}_{\mathrm{s}} \in \mathbb{C}^{L_{\mathrm{s}} \times 1}$ denotes the quantized downlink DPGI acquired at BS in (\ref{g_s_qua_est}).
	For the scenario of $L_{\mathrm{s}} = L$, NMSE is given by $\mathbb{E} \left[ \| \tilde{\mathbf{g}}_{\mathrm{s}} - \mathbf{g} \|^2 / \|\mathbf{g}\|^2 \right]$.
	In addition, the NMSEs of LS estimator and MMSE estimator are respectively expressed as $\mathbb{E} \left[ \| \hat{\mathbf{g}}_{\mathrm{LS}} - \mathbf{g} \|^2 / \|\mathbf{g}\|^2 \right]$ and $\mathbb{E} \left[ \| \hat{\mathbf{g}}_{\mathrm{MMSE}} - \mathbf{g} \|^2 / \|\mathbf{g}\|^2 \right]$ \cite{Xing_MM_MIMO}.$\footnote{Different from our proposed estimator, UE side cannot perform the PGI/DPGI estimation for LS estimator and MMSE estimator because PAI is not available at UE, which means that the gains of different paths cannot be obtained and separated at UE. Therefore, received signal vector $\mathbf{y} \in \mathbb{C}^{T \times 1}$ has to be fed back in uplink and the whole PGI (i.e., gains of all $L$ cascaded paths) will be estimated at BS.}$

	Thanks to the joint beamforming design in (\ref{act_ce}) and (\ref{pass_ce}), the proposed DPGI/PGI estimator outperforms the LS and MMSE estimators at a relatively low PNR region, as shown in Fig.~\ref{fig2}.
	Compared with the proposed scheme which requires only $T = L_{\mathrm{s}}$ pilot signals, much more pilot signals are required for both LS estimator and MMSE estimator to achieve a comparable accuracy of estimation.
	Besides, the equivalent noise of proposed scheme in (\ref{y_ce_4}) consists of two parts: the channel noise $n(t)$ and interference from other $(L - 1)$ paths $\sum_{l \neq \lambda_{\mathrm{s},t}}^{L} \kappa_{l}(t) g_l$.
	The estimation accuracy of proposed scheme is mainly affected by the channel noise at first, so NMSE decreases slowly at a relatively low PNR region.
	Then, the interference of other paths dominates the equivalent noise around when PNR $>$ -5dB, so the NMSE curves of proposed estimator tend to be stable.
	In addition, the NMSE performance of MMSE estimator is better than that of LS estimator, so we will only adopt MMSE estimator (with $T = 24$) in the following simulations.

	\begin{figure}[t]
		\center{\includegraphics[width=0.5\textwidth]{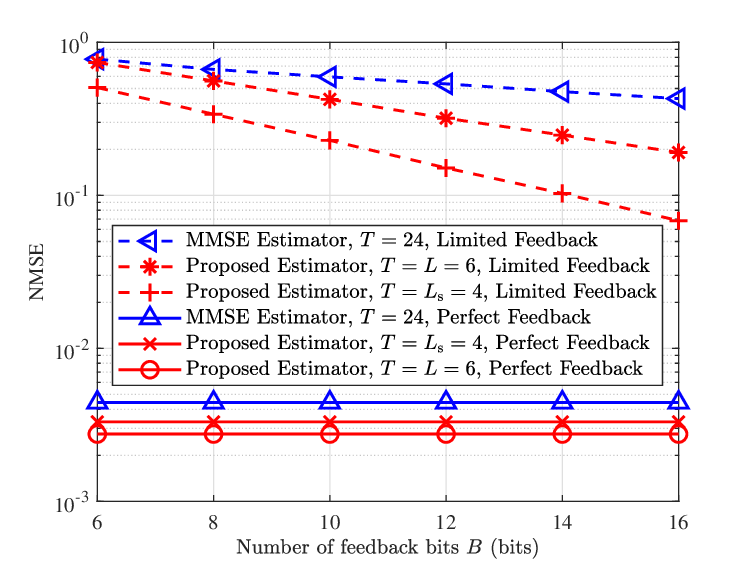}}
		\caption{NMSE performance against number of feedback bits $B$ when $N_{\mathrm{B}} = N_{\mathrm{B,v}} \times N_{\mathrm{B,h}} = 4 \times 4$, $N_{\mathrm{R}} = N_{\mathrm{R,v}} \times N_{\mathrm{R,h}} = 16 \times 16$, PNR $= 0$ dB, $L = L_{\mathrm{RB}} \times L_{\mathrm{RU}} = 2 \times 3$.}
		\label{fig3}
	\end{figure}

	Fig.~\ref{fig3} shows the NMSE performance against the number of feedback bits $B$ when PNR $= 0$ dB. 
	The solid lines refer to the case using perfect feedback ($B \to \infty$) and dashed lines refer to the case with limited feedback.
	In terms of the NMSE performance, the proposed estimator outperforms MMSE estimator with limited $B$ for two main reasons.
	Firstly, the minimal achievable NMSE of proposed estimator (corresponding to solid lines when $B \to \infty$) is lower than that of MMSE estimator.
	Secondly, the pilot length of the proposed estimator (i.e., $T$, equal to the length of feedback vector) is smaller, so the inevitable quantization error caused by the RVQ feedback of proposed estimator is smaller than that of MMSE estimator with fixed $B$.
	Thus, the achievable NMSE of our proposed estimator with $T = L_{\mathrm{s}} = 4$ is lower than that of our proposed estimator with $T = L = 6$ for limited feedback scenario.

	\begin{figure}[t]
		\center{\includegraphics[width=0.5\textwidth]{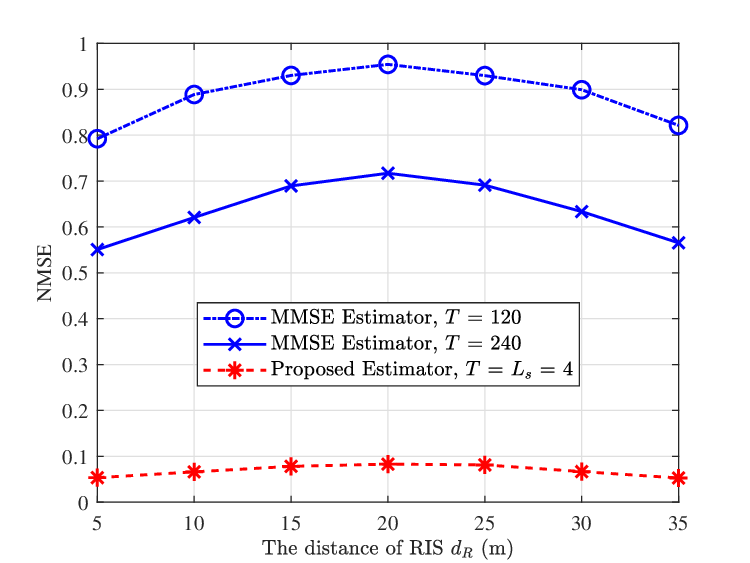}}
		\caption{NMSE performance against the distance of RIS $d_{\mathrm{R}}$ when $N_{\mathrm{B}} = N_{\mathrm{B,v}} \times N_{\mathrm{B,h}} = 4 \times 4$, $N_{\mathrm{R}} = N_{\mathrm{R,v}} \times N_{\mathrm{R,h}} = 16 \times 16$, PNR $= 30$ dB, $L = L_{\mathrm{RB}} \times L_{\mathrm{RU}} = 2 \times 3$.}
		\label{fig4}
	\end{figure}

	The impact of RIS's location on the NMSE performance is considered in Fig.~\ref{fig4}, where we assume that the coordinates of the BS, the RIS, and the UE are $(0,0)$ m, $(d_{\mathrm{R}},10)$ m, and $(40,0)$ m, respectively.
	Then, the distance between the BS and the RIS and that between the RIS and the UE can be given as $d_{\mathrm{RB}} = \sqrt{d_{\mathrm{R}}^2 + 10^2}$ and $d_{\mathrm{RU}} = \sqrt{(40 - d_{\mathrm{R}})^2 + 10^2}$, respectively.
	To characterize the impact of distance-dependent path loss and antenna gain on the two cascaded channels, we multiply the $\mathbf{H}_{\mathrm{RB}}$ in (\ref{H_RB}) and the $\mathbf{h}_{\mathrm{RU}}$ in (\ref{h_RU}) by $\sqrt{\rho_{\mathrm{RB}}}$ and $\sqrt{\rho_{\mathrm{RU}}}$, respectively, where we have $\rho_{\mathrm{RB}} = -30 - 22 \mathrm{log}_{10} d_{\mathrm{RB}} + G_{\mathrm{B}}$ dB ($\rho_{\mathrm{RU}} = -30 - 22 \mathrm{log}_{10} d_{\mathrm{RU}} + G_{\mathrm{U}}$ dB) with $G_{\mathrm{B}} = 28$ dBi ($G_{\mathrm{U}} = 25$ dBi) denoting the transmitting (receiving) antenna gain \cite{RIS_location_FengCH_IOT_2022,RIS_location_VTC_2022}.
	It is observed that as the distance of RIS $d_{\mathrm{R}}$ increases, the NMSE increases and then reaches the peak at $d_{\mathrm{R}} = 20$ m, then decreases.
	This is because the total path loss is the product of the path losses of two cascaded channels.
	When the location of RIS is close to the midpoint between the BS and the UE, the distance-dependent path loss is maximized. 
	Therefore, to improve the NMSE performance, it would be good to place the RIS near the BS or UE.
	In addition, as shown in Fig.~\ref{fig4}, our proposed estimator exhibits less sensitivity to the distance-dependent path loss in comparison to the MMSE estimator. 
	This is because that the accuracy of our proposed estimator is mainly affected by the interference from other paths (corresponding to the $\sum_{l \neq \lambda_{\mathrm{s},t}}^{L} \kappa_{l}(t) g_l$ in (\ref{y_ce_4})), rather than the channel noise which significantly impairs the NMSE performance when the path loss is high.

    \subsection{Performance of Path Selection and Beamforming}\label{S4.2}

	In this subsection, we compare the proposed path selection and beamforming scheme with conventional beamforming schemes.
	To form an effective comparison, the alternating optimization proposed in \cite{BF_RIS_JointA&P_ZhangRui_TWC_2019} is also simulated, which can be used for the design of active beamformer $\mathbf{f}_{\mathrm{t}}$ and passive beamformer $\boldsymbol{\psi}$.
	Specifically, by defining
	\begin{align}\label{H_sub}
		\mathbf{H} = \sqrt{\frac{N_{\mathrm{B}} N_{\mathrm{R}}^{2}}{L}} \tilde{\mathbf{A}}_{\mathrm{R}} \operatorname{diag}(\mathbf{g}) \tilde{\mathbf{A}}^{\rm{H}}_{\mathrm{B}} \in \mathbb{C}^{N_{\mathrm{R}} \times N_{\mathrm{B}}}
	\end{align}
	according to (\ref{y_ce_1}), the cascaded channel is re-expressed as $\mathbf{h}^{\rm{H}} = \boldsymbol{\psi}^{\rm{T}} \mathbf{H} \in \mathbb{C}^{1 \times N_{\mathrm{B}}}$, and the objective function $\mathbb{E}\left[\left|\mathbf{h}^{\rm{H}} \mathbf{f}_{\mathrm{t}}\right|^{2}\right]$ is transformed into $\mathbb{E}\left[\left|\boldsymbol{\psi}^{\rm{T}} \mathbf{H} \mathbf{f}_{\mathrm{t}}\right|^{2}\right]$.
	Then, active beamformer $\mathbf{f}_{\mathrm{t}}$ and passive beamformer $\boldsymbol{\psi}$ are alternatively optimized by $\mathbf{f}_{\mathrm{t}} = \frac{\mathbf{H}^{\rm{H}} \boldsymbol{\psi}^{*}} {\| \mathbf{H}^{\rm{H}} \boldsymbol{\psi}^{*} \|} \in \mathbb{C}^{N_{\mathrm{B}} \times 1}$ and $\boldsymbol{\psi} = \frac{\mathbf{H}^{*} \mathbf{f}^{*}_{\mathrm{t}}} {|\mathbf{H} \mathbf{f}_{\mathrm{t}}|} \in \mathbb{C}^{N_{\mathrm{R}} \times 1}$ until $\left|\boldsymbol{\psi}^{\rm{T}} \mathbf{H} \mathbf{f}_{\mathrm{t}}\right|^{2}$ converges \cite{BF_RIS_JointA&P_ZhangRui_TWC_2019}.
	For the case where the whole PGI (i.e., $\mathbf{g} \in \mathbb{C}^{L \times 1}$) can be obtained, the quantized estimate of $\mathbf{g} \in \mathbb{C}^{L \times 1}$ can be brought into (\ref{H_sub}) for subsequent alternating optimizations.
	In addition, for the case of $T = L_{\mathrm{s}} < L$, a scheme named `Beamforming in \cite{BF_RIS_JointA&P_ZhangRui_TWC_2019} with Partial Random PGI' is also provided to form an effective comparison with our proposed DPGI-based design of active beamformer and passive beamformer.
	Specifically, for the PGI (i.e., $\mathbf{g} \in \mathbb{C}^{L \times 1}$) brought into (\ref{H_sub}), we assign the elements corresponding to selected dominant paths in $\mathbf{g} \in \mathbb{C}^{L \times 1}$ to the value of quantized downlink DPGI $\tilde{\mathbf{g}}_{\mathrm{s}} \in \mathbb{C}^{L_{\mathrm{s}} \times 1}$ in (\ref{g_s_qua_est}), and the remaining elements in $\mathbf{g} \in \mathbb{C}^{L \times 1}$ with dimension of $L_{\mathrm{r}} \times 1$ are assigned to random values. Then, the alternating iteration proposed in \cite{BF_RIS_JointA&P_ZhangRui_TWC_2019} is adopted for subsequent active and passive beamforming design.

	\begin{figure}[t]
		\center{\includegraphics[width=0.5\textwidth]{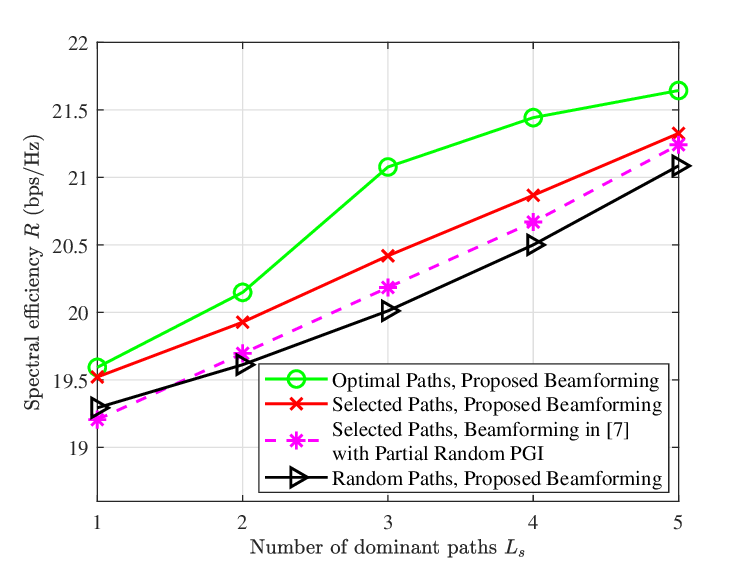}}
		\caption{Spectral efficiency $R$ for downlink data transmission against number of dominant paths $L_\mathrm{s}$ when $N_{\mathrm{B}} = N_{\mathrm{B,v}} \times N_{\mathrm{B,h}} = 4 \times 4$, $N_{\mathrm{R}} = N_{\mathrm{R,v}} \times N_{\mathrm{R,h}} = 16 \times 16$, DNR $= 10$ dB, $L = L_{\mathrm{RB}} \times L_{\mathrm{RU}} = 2 \times 3$.}
		\label{fig5}
	\end{figure}

	Spectral efficiency $R$ for the downlink data transmission against the number of dominant paths $L_\mathrm{s}$ is shown in Fig.~\ref{fig5}, where the ideal estimation and the perfect feedback of DPGI are temporarily assumed here (i.e., the accurate DPGI is available at BS, and the scenario of non-ideal estimation and limited feedback is considered in the following simulations).
	We observe that although the achievable $R$ of our proposed path selection (denoted as `Selected Paths') is smaller than that of optimal path selection (obtained by exhaustive search, denoted as `Optimal Paths'), our proposed path selection significantly outperforms the random path selection (denoted as `Random Paths').
	In addition, when dominant paths are selected according to our proposed scheme, compared with the beamforming in \cite{BF_RIS_JointA&P_ZhangRui_TWC_2019} with partial random PGI, our proposed scheme in Subsection \ref{S3.3} achieves higher spectral efficiency $R$.

	\begin{figure}[t]
		\center{\includegraphics[width=0.5\textwidth]{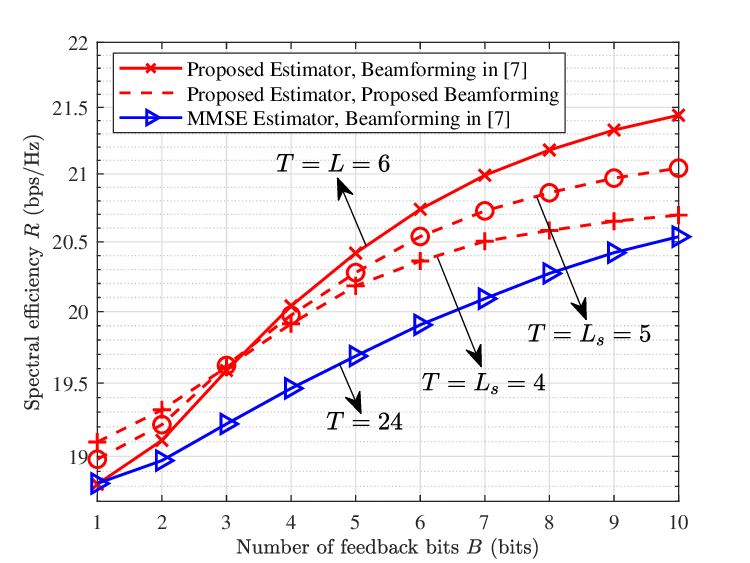}}
		\caption{Spectral efficiency $R$ for downlink data transmission against number of feedback bits $B$ when $N_{\mathrm{B}} = N_{\mathrm{B,v}} \times N_{\mathrm{B,h}} = 4 \times 4$, $N_{\mathrm{R}} = N_{\mathrm{R,v}} \times N_{\mathrm{R,h}} = 16 \times 16$, PNR $= -15$ dB, DNR $= 10$ dB, $L = L_{\mathrm{RB}} \times L_{\mathrm{RU}} = 2 \times 3$.}
		\label{fig6}
	\end{figure}

	As shown in Fig.~\ref{fig6}, when the quantized estimate of PGI is used for the beamforming design proposed in \cite{BF_RIS_JointA&P_ZhangRui_TWC_2019} (corresponding to lines with $T = L = 6$ and $T = 24$), higher spectral efficiency $R$ can be achieved by adopting our proposed estimator rather than adopting MMSE estimator due to the estimation accuracy and small feedback overhead of the proposed scheme.
	Besides, when the quantized estimate of DPGI is used for our proposed beamforming scheme (corresponding to lines with $T = L_{\mathrm{s}} = 5$ and $T = L_{\mathrm{s}} = 4$), although there is a certain performance loss in the achievable spectral efficiency $R$ compared with aforementioned scheme using the whole PGI (corresponding to line with $T = L = 6$), higher spectral efficiency $R$ can be achieved around when $B < 4$.
	This is because when the number of feedback bits $B$ is fixed, the accuracy of DPGI fed back to the BS increases with the decrease of feedback length $T$, which further leads to a corresponding increase of spectral efficiency $R$.
	This clearly demonstrates the significance and superiority of our proposed path selection strategy.

	\begin{figure}[t]
		\center{\includegraphics[width=0.5\textwidth]{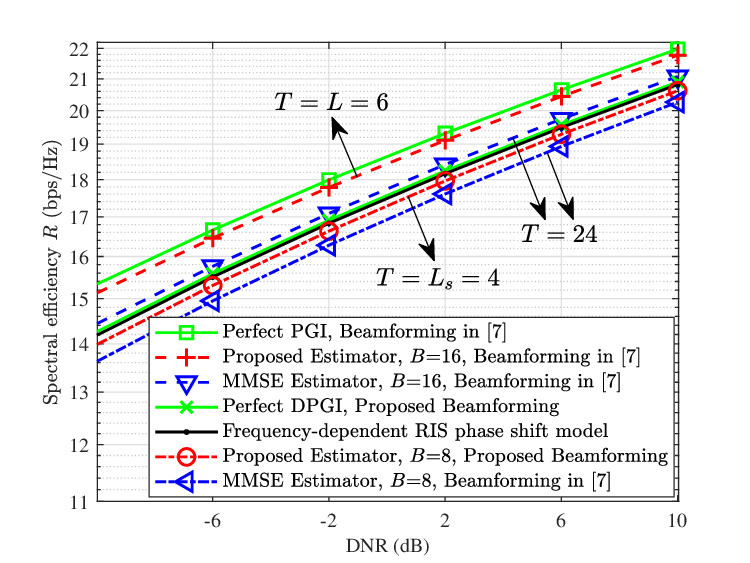}}
		\caption{Spectral efficiency $R$ for downlink data transmission against DNR when $N_{\mathrm{B}} = N_{\mathrm{B,v}} \times N_{\mathrm{B,h}} = 4 \times 4$, $N_{\mathrm{R}} = N_{\mathrm{R,v}} \times N_{\mathrm{R,h}} = 16 \times 16$, PNR $= -15$ dB, $L = L_{\mathrm{RB}} \times L_{\mathrm{RU}} = 2 \times 3$, $L_{\mathrm{s}} = 4$.}
		\label{fig7}
	\end{figure}

	All the achievable spectral efficiency $R$ under different settings increases exponentially with DNR as shown in Fig.~\ref{fig7}.
	For the scenario where the whole PGI is exploited, the maximum $R$ can be achieved when the perfect PGI (ideal PGI estimation and perfect feedback) is used for beamforming in \cite{BF_RIS_JointA&P_ZhangRui_TWC_2019}.
	In this case, the spectral efficiency gap can be controlled within 0.3 bps/Hz by exploiting our proposed estimator and a 16-bit RVQ codebook (corresponding to line with $T = L = 6$).
	Besides, by exploiting our proposed path selection and beamforming scheme, the maximum $R$ is achieved with perfect DPGI (ideal DPGI estimation and perfect feedback).
	In this case, by adopting our proposed estimator and a 8-bit RVQ codebook (corresponding to line with $T = L_{\mathrm{s}} = 4$), the spectral efficiency gap is also controlled within 0.3 bps/Hz.
	The performance degradation of our proposed DPGI-based scheme can be observed by comparing the two lines labeled as `MMSE Estimator, $B = 16$, Beamforming in [7]' and `Proposed Estimator, $B = 8$, Proposed Beamforming'.
	The gap of the achievable spectral efficiency $R$ between these two lines is less than 0.4 bps/Hz.
	We also consider the impact of frequency-dependent RIS phase shifts with an angle difference $\phi_{\mathrm{bias}} = 10^{\circ}$ between the downlink and uplink channels.$\footnote{The angle difference between the uplink and downlink frequency-dependent RIS phase shifts introduces disparities in matrix $\mathbf{B}_{l}$ in (\ref{B_defi_2}) for the uplink and downlink channels. Fortunately, this angle difference can be calculated (or experimentally measured) for a given communication system \cite{FreqdepRIS_PScal_mmWave_2023, FreqdepRIS_PSmeasure_mmWave_2023}.
	It can be observed from the Fig 1.(b) in \cite{FreqdepRIS_PSmeasure_mmWave_2023} that the value of $\phi_{\mathrm{bias}}$ corresponding to our considered downlink and uplink carrier frequencies is less than 3 degrees.
	By compensating for the angle difference, one can ensure the angle reciprocity between the uplink and downlink channels.}$
	The performance degradation caused by frequency-dependent RIS phase shifts can be observed by comparing the two lines labeled as `Perfect DPGI, Proposed Beamforming' and `Frequency-dependent RIS phase shift model'. 
	The gap of the achievable spectral efficiency $R$ between these two lines is less than 0.1 bps/Hz.
	In addition, it can be observed that the spectral efficiency performance of our proposed estimator is much better than the spectral efficiency performance of the conventional MMSE estimator.

	\section{Conclusions}\label{S5}

	In this paper, we proposed a path selection based feedback reduction and partial CSI-based beamforming scheme for the FDD RIS-assisted systems.
	Specifically, downlink PAI was first acquired at BS via the angle reciprocity.
	We also proposed a path selection strategy by removing the path with minimal contribution to spectral efficiency sequentially, during which the active and passive beamformers are alternatively optimized.
	Furthermore, we proposed a DPGI estimation and feedback scheme, where both the length of downlink pilot signals and feedback vector are reduced to the number of selected dominant paths.
	Finally, we further improved the spectral efficiency of downlink data transmission by updating the active and passive beamformers based on the quantized estimate of DPGI.
	From the numerical experiments, we could observe the performance gain of proposed algorithms over the conventional schemes.
	We plan to consider more relevant cases in our future work (e.g., the UE equipped with multiple antennas and the multi-user case in FDD RIS-assisted communication systems).

	\appendices
	\section{Proof of Lemma 1}

	According to the channel decomposition in (\ref{h_AsArg}), the objective function in (\ref{goal_inial}) can be given by \cite{FB_Pathsel_Shim_TWC_2020}
	\begin{align}
		\mathbb{E}\left[\left|\mathbf{h}^{\rm{H}} \mathbf{f}_{\mathrm{t}}\right|^{2}\right] &  = \mathbb{E}\left[\left| \mathbf{g}^{\rm{T}}_{\mathrm{s}} \mathbf{A}^{\rm{H}}_{\mathrm{s}} \mathbf{V} \mathbf{g}^{*}_{\mathrm{s}} + \mathbf{g}^{\rm{T}}_{\mathrm{r}} \mathbf{A}^{\rm{H}}_{\mathrm{r}} \mathbf{V} \mathbf{g}^{*}_{\mathrm{s}} \right|^{2}\right] \label{goal_decom_1} \\
		& = \mathbb{E}\left[\left| \mathbf{g}^{\rm{T}}_{\mathrm{s}} \mathbf{A}^{\rm{H}}_{\mathrm{s}} \mathbf{V} \mathbf{g}^{*}_{\mathrm{s}} \right|^{2}\right] + \mathbb{E}\left[\left| \mathbf{g}^{\rm{T}}_{\mathrm{r}} \mathbf{A}^{\rm{H}}_{\mathrm{r}} \mathbf{V} \mathbf{g}^{*}_{\mathrm{s}} \right|^{2}\right] \label{goal_decom_2} \\
		&  = \mathbb{E}\left[\left| \mathbf{g}^{\rm{T}}_{\mathrm{s}} \mathbf{W}_{\mathrm{s}} \mathbf{g}^{*}_{\mathrm{s}} \right|^{2}\right] + \mathbb{E}\left[\left| \mathbf{g}^{\rm{T}}_{\mathrm{r}} \mathbf{W}_{\mathrm{r}} \mathbf{g}^{*}_{\mathrm{s}} \right|^{2}\right], \label{goal_decom_3}
	\end{align}	
	where (\ref{goal_decom_2}) is derived according to $\mathbb{E}\left[ \mathbf{g}^{*}_{\mathrm{s}} \mathbf{g}^{\rm{T}}_{\mathrm{r}} \right] = \mathbf{0}_{L_{\mathrm{s}} \times L_{\mathrm{r}}}$, since $\mathbb{E}\left[g^{*}_{\mathrm{s}, i} g_{\mathrm{r}, j}\right] = \mathbb{E}\left[ \alpha^{*}_{p} \beta^{*}_{q} \alpha_{m} \beta_{n} \right] = \mathbb{E}\left[\alpha^{*}_{p} \alpha_{m}\right] \mathbb{E}\left[\beta^{*}_{q} \beta_{n}\right] = 0$ for $\forall i = (q-1) L_{\mathrm{RB}} + p \in \Lambda_{\mathrm{s}}$ and $\forall j = (n-1) L_{\mathrm{RB}} + m \in \Lambda_{\mathrm{r}}$ satisfying $\{p, q\} \neq\{m, n\}$.
	Besides, $\mathbf{W}_{\mathrm{s}} = \mathbf{A}_{\mathrm{s}}^{\rm{H}} \mathbf{V} \in \mathbb{C}^{L_{\mathrm{s}} \times L_{\mathrm{s}}}$ and $\mathbf{W}_{\mathrm{r}} = \mathbf{A}_{\mathrm{r}}^{\rm{H}} \mathbf{V} \in \mathbb{C}^{L_{\mathrm{r}} \times L_{\mathrm{s}}}$ are defined in (\ref{goal_decom_3}) for simplicity. Here we respectively redefine variables $i$ and $j$ corresponding to values of the $l_{i}$-th element and the $l_{j}$-th element in $\Lambda_{\mathrm{s}}$, where $1 \leq i,j \leq L$, $1 \leq l_i,l_j \leq L_{\mathrm{s}}$.
	Variables $i$ and $j$ also denote the indices of selected dominant cascaded paths, where we assume $i = (q-1) L_{\mathrm{RB}} + p$ and $j = (n-1) L_{\mathrm{RB}} + m$.
	The first term in (\ref{goal_decom_3}) is further given by \cite{FB_Pathsel_Shim_TWC_2020}
	\begin{align}\label{W_s}
		\nonumber & \mathbb{E} \! \left[\left| \mathbf{g}_{\mathrm{s}}^{\rm{T}} \mathbf{W}_{\mathrm{s}} \mathbf{g}^{*}_{\mathrm{s}}\right|^{2}\right] \\
		= & \ \mathbb{E} \! \left[\left| \sum\limits_{i \in \Lambda_{\mathrm{s}}} \! \mathbf{W}^{l_i, l_i}_{\mathrm{s}} |g_{\mathrm{s}, l_i}|^{2} \! + \! \sum\limits_{i \neq j} \mathbf{W}^{l_i, l_j}_{\mathrm{s}} g_{\mathrm{s}, l_i} g^{*}_{\mathrm{s}, l_j} \right|^{2}\right] \\
		= \! & \ \mathbb{E} \! \left[\left|\sum\limits_{i \in \Lambda_{\mathrm{s}}} \! \mathbf{W}^{l_i, l_i}_{\mathrm{s}} \! \left| g_{\mathrm{s}, l_i}\right|^{2}\right|^{2}\right] \!\! + \! \mathbb{E} \! \left[\left|\sum\limits_{i \neq j} \! \mathbf{W}^{l_i, l_j}_{\mathrm{s}} g_{\mathrm{s}, l_i} g^{*}_{\mathrm{s}, l_j}\right|^{2}\right] \label{W_s_3} \! \\
		\nonumber = & \ \sum\limits_{i \in \Lambda_{\mathrm{s}}} \left|\mathbf{W}^{l_i, l_i}_{\mathrm{s}}\right|^{2} \mathbb{E}\left[\left|g_{\mathrm{s}, l_i}\right|^{4}\right] + \sum\limits_{i \neq j} \left(\mathbf{W}^{l_i, l_i}_{\mathrm{s}}\right)^{*} \mathbf{W}^{l_j, l_j}_{\mathrm{s}} \\
		& \times \! \mathbb{E} \! \left[\left|g_{\mathrm{s}, l_i}\right|^{2} \! \left|g_{\mathrm{s}, l_j}\right|^{2}\right] \! + \! \sum\limits_{i \neq j} \! \left|\mathbf{W}^{l_i, l_j}_{\mathrm{s}}\right|^{2} \mathbb{E} \! \left[\left|g_{\mathrm{s}, l_i}\right|^{2} \! \left|g_{\mathrm{s}, l_j}\right|^{2}\right] \! , \! \label{W_s_4}
	\end{align}	
	where (\ref{W_s_3}) is due to $\mathbb{E}\left[ g_{\mathrm{s}, l_i} g^{*}_{\mathrm{s}, l_j} \right] = 0$ for $\forall \ l_i \neq l_j$.
	Then, we have
	\begin{align}\label{g_s_4_same}
		\nonumber & \mathbb{E} \! \left[\left|g_{\mathrm{s}, l_i}\right|^{4}\right] \! = \! \mathbb{E}\! \left[ \left|\alpha_{p}\right|^{4} \right] \! \mathbb{E} \! \left[ \left|\beta_{q}\right|^{4} \right] \! = \! \left( \! \left( \! \mathbb{E} \! \left[\left|\alpha_{p}\right|^{2}\right]\right)^{\! 2} \!\! + \! \mathbb{D} \! \left[\left|\alpha_{p}\right|^{2}\right] \! \right) \\
		& \times \! \left( \! \left( \! \mathbb{E} \! \left[\left|\beta_{q}\right|^{2}\right]\right)^{\! 2} \!\! + \! \mathbb{D} \! \left[\left|\beta_{q}\right|^{2} \right]\right) \! = \! (1^2 + 1) \! \times \! (1^2 + 1) \! = \! 4, \!
	\end{align}	
	and
	\begin{align}\label{g_s_4times}
		\mathbb{E}\left[\left|g_{\mathrm{s}, l_i}\right|^{2}\left|g_{\mathrm{s}, l_j}\right|^{2}\right] = \mathbb{E}\left[ \left|\alpha_{p}\right|^{2} \left|\alpha_{m}\right|^{2} \left|\beta_{q}\right|^{2} \left|\beta_{n}\right|^{2} \right],
	\end{align}		
	where
	\begin{align}
		\nonumber & \mathbb{E}\left[\left|\alpha_{p}\right|^{2}\left|\alpha_{m}\right|^{2}\right] \\
		= & \begin{cases} \! \mathbb{E} \! \left[\left|\alpha_{p}\right|^{4}\right] \! = \! \left( \! \mathbb{E} \! \left[\left|\alpha_{p}\right|^{2}\right]\right)^{\! 2} \!\! + \! \mathbb{D} \! \left[\left|\alpha_{p}\right|^{2}\right] \! = \! 2, \!\! & \!\! \mathrm{\! for} \ p \! = \! m \\
		\! \mathbb{E} \! \left[\left|\alpha_{p}\right|^{2}\right] \! \mathbb{E} \! \left[\left|\alpha_{m}\right|^{2}\right]=1, \!\! & \!\! \mathrm{\! for}\ p \! \neq \! m \end{cases} \! , \! \label{E_cal_classification_1} \\
		\nonumber & \mathbb{E}\left[\left|\beta_{q}\right|^{2}\left|\beta_{n}\right|^{2}\right] \\
		= & \begin{cases} \! \mathbb{E} \! \left[\left|\beta_{q}\right|^{4}\right] \! = \! \left( \! \mathbb{E} \! \left[\left|\beta_{q}\right|^{2}\right]\right)^{\! 2} \!\! + \! \mathbb{D} \!  \left[\left|\beta_{q}\right|^{2}\right] \! = \! 2, \!\! & \!\! \mathrm{\! for}\ q \! = \! n \\ \! \mathbb{E} \! \left[\left|\beta_{q}\right|^{2}\right] \! \mathbb{E} \! \left[\left|\beta_{n}\right|^{2}\right]=1, \!\! & \!\! \mathrm{\! for}\ q \! \neq \! n \end{cases} \! . \! \label{E_cal_classification_2}
	\end{align}	
	For a given cascaded path index $i = (q-1) L_{\mathrm{RB}} + p$ with a combination $\{p, q\}$, the remaining $(L - 1)$ optional indices of $j = (n-1) L_{\mathrm{RB}} + m$ with a combination $\{m, n\}$ can be divided into three types:
	\textit{\romannumeral1)} The number of optional indices satisfying $(p \neq m) \cap(q = n)$ is equal to $\left( L_{\mathrm{RB}} - 1 \right)$;
	\textit{\romannumeral2)} The number of optional indices satisfying $(p = m) \cap(q \neq n)$ is equal to $\left( L_{\mathrm{RU}} - 1 \right)$;
	\textit{\romannumeral3)} The number of optional indices satisfying $(p \neq m) \cap(q \neq n)$ is equal to $\left( L - L_{\mathrm{RB}} - L_{\mathrm{RU}} + 1 \right)$.
	Therefore, based on (\ref{E_cal_classification_1}) and (\ref{E_cal_classification_2}), the closed-form expression of (\ref{g_s_4times}) is given by	
	\begin{align}\label{g_s_4_final}
		\nonumber & \mathbb{E} \left[ \left|\alpha_{p}\right|^{2} \left|\alpha_{m}\right|^{2} \left|\beta_{q}\right|^{2} \left|\beta_{n}\right|^{2} \right] \\
		\nonumber = & \frac{2 \left( L_{\mathrm{RB}} - 1 \right) + 2 \left( L_{\mathrm{RU}} - 1 \right) + (L - L_{\mathrm{RB}} - L_{\mathrm{RU}} + 1)}{L_{\mathrm{RB}} L_{\mathrm{RU}}-1} \\
		= & \frac{L + L_{\mathrm{RB}} + L_{\mathrm{RU}} - 3}{L-1}.
	\end{align}	
	By substituting (\ref{g_s_4_same}) and (\ref{g_s_4_final}) into (\ref{W_s_4}), we get
	\begin{align}\label{W_s_cal}
		\nonumber & \mathbb{E}\left[\left| \mathbf{g}^{\rm{T}}_{\mathrm{s}} \mathbf{W}_{\mathrm{s}} \mathbf{g}^{*}_{\mathrm{s}} \right|^{2}\right] \\
		\nonumber = & \ 4 \! \sum\limits_{l_{\mathrm{s}} = 1}^{L_{\mathrm{s}}} \! \left|\mathbf{W}^{l_s, l_s}_{\mathrm{s}}\right|^{2} \!\! + \! Q \!\! \sum\limits_{l_i \neq l_j}^{L_{\mathrm{s}}} \!\! \left(\mathbf{W}^{l_i, l_i}_{\mathrm{s}} \! \right)^{\! *} \! \mathbf{W}^{l_j, l_j}_{\mathrm{s}} \! + \! Q \!\! \sum\limits_{l_i \neq l_j}^{L_{\mathrm{s}}} \!\! \left|\mathbf{W}^{l_i, l_j}_{\mathrm{s}}\right|^{2} \! \\
		= & \ (4 - 2 Q) \left\|\operatorname{diag} \left(\mathbf{W}_{\mathrm{s}}\right)\right\|^{2} \! + \! Q |\operatorname{tr}(\mathbf{W}_{\mathrm{s}})|^{2} \! + \! Q \|\mathbf{W}_{\mathrm{s}}\|_{\rm{F}}^{2},
	\end{align}	
	where we define $Q = \frac{L+L_{\mathrm{RB}}+L_{\mathrm{RU}}-3}{L-1}$.
	Similarly, the second term in (\ref{goal_decom_3}) is calculated as
	\begin{align}\label{W_r_cal}
		\mathbb{E}\left[\left| \mathbf{g}^{\rm{T}}_{\mathrm{r}} \mathbf{W}_{\mathrm{r}} \mathbf{g}^{*}_{\mathrm{s}} \right|^{2}\right] = Q \sum\limits_{l_{\mathrm{r}} = 1}^{L_{\mathrm{r}}} \sum\limits_{l_{\mathrm{s}} = 1}^{L_{\mathrm{s}}} \left| \mathbf{W}^{l_{\mathrm{r}}, l_{\mathrm{s}}}_{\mathrm{r}} \right|^{2} = Q \left\| \mathbf{W}_{\mathrm{r}} \right\|^{2}_{\rm{F}}.
	\end{align}	
	Finally, the expression of $\mathbb{E}\left[\left|\mathbf{h}^{\rm{H}} \mathbf{f}_{\mathrm{t}}\right|^{2}\right]$ in Lemma 1 is obtained by combining (\ref{W_s_cal}) and (\ref{W_r_cal}).

	\section{Proof of Lemma 2}

	For a given $\mathbf{g}_{\mathrm{s}} \in \mathbb{C}^{L_{\mathrm{s}} \times 1}$, the following equalities hold  \cite{FB_Pathsel_Shim_TWC_2020,FB_Energy_Shim_TCOM_2021}
	\begin{align}
		\mathbb{E}\left[\left|\mathbf{h}^{\rm{H}} \mathbf{f}_{\mathrm{t}}\right|^{2}\right] = & \ \mathbb{E}\left[\left|\mathbf{g}_{\mathrm{r}}^{\rm{T} } \mathbf{A}_{\mathrm{r}}^{\rm{H}} \mathbf{f}_{\mathrm{t}}\right|^{2}\right] + \left|\mathbf{g}_{\mathrm{s}}^{\rm{T}} \mathbf{A}_{\mathrm{s}}^{\rm{H}} \mathbf{f}_{\mathrm{t}}\right|^{2} \label{goal_update_1} \\
		= & \ \mathbb{E}\left[ \mathbf{f}_{\mathrm{t}}^{\rm{H}} \mathbf{A}_{\mathrm{r}} \mathbf{g}^{*}_{\mathrm{r}} \mathbf{g}^{\rm{T}}_{\mathrm{r}} \mathbf{A}_{\mathrm{r}}^{\rm{H}} \mathbf{f}_{\mathrm{t}} \right] + \left|\mathbf{g}_{\mathrm{s}}^{\rm{T}} \mathbf{A}_{\mathrm{s}}^{\rm{H}} \mathbf{f}_{\mathrm{t}} \right|^{2} \\
		= & \ \mathbf{f}_{\mathrm{t}}^{\rm{H}} \mathbf{A}_{\mathrm{r}} \mathbb{E}\left[ \mathbf{g}^{*}_{\mathrm{r}} \mathbf{g}^{\rm{T}}_{\mathrm{r}} \right] \mathbf{A}_{\mathrm{r}}^{\rm{H}} \mathbf{f}_{\mathrm{t}} + \left|\mathbf{g}_{\mathrm{s}}^{\rm{T}} \mathbf{A}_{\mathrm{s}}^{\rm{H}} \mathbf{f}_{\mathrm{t}} \right|^{2},  \label{goal_update_3}
	\end{align}	
	where (\ref{goal_update_1}) is derived according to the channel decomposition in (\ref{h_AsArg}).
	Here we redefine variables $i = (q-1) L_{\mathrm{RB}} + p \in \Lambda_{\mathrm{r}}$ and $j = (n-1) L_{\mathrm{RB}} + m \in \Lambda_{\mathrm{r}}$.
	Then, we have
	\begin{align}
		\nonumber \mathbb{E}\left[ g^{*}_{\mathrm{r}, i} g_{\mathrm{r},j} \right] = & \ \mathbb{E}\left[ \alpha_{p}^{*} \alpha_{m} \right] \mathbb{E}\left[ \beta_{q}^{*} \beta_{n} \right] \\
		= & \begin{cases} 1, & \mathrm{for} \ i=j, \{p, q\} =\{m, n\} \\  0, & \mathrm{for} \ i \neq j, \{p, q\} \neq \{m, n\} \end{cases},
	\end{align}	
	and further we get $\mathbb{E}\left[ \mathbf{g}^{*}_{\mathrm{r}} \mathbf{g}^{\rm{T}}_{\mathrm{r}} \right] = \mathbf{I}_{L_{\mathrm{r}}}$.
	Finally, the objective function in (\ref{goal_update_1}) can be formulated as
	\begin{align}
		\nonumber \mathbb{E}\left[\left|\mathbf{h}^{\rm{H}} \mathbf{f}_{\mathrm{t}}\right|^{2}\right] = & \ \mathbf{f}_{\mathrm{t}}^{\rm{H}} \mathbf{A}_{\mathrm{r}} \mathbf{A}_{\mathrm{r}}^{\rm{H}} \mathbf{f}_{\mathrm{t}} + \left|\mathbf{g}_{\mathrm{s}}^{\rm{T}} \mathbf{A}_{\mathrm{s}}^{\rm{H}} \mathbf{f}_{\mathrm{t}} \right|^{2} \\
		= & \left\|\mathbf{A}_{\mathrm{r}}^{\rm{H}} \mathbf{f}_{\mathrm{t}} \right\|^{2} + \left|\mathbf{g}_{\mathrm{s}}^{\rm{T}} \mathbf{A}_{\mathrm{s}}^{\rm{H}} \mathbf{f}_{\mathrm{t}} \right|^{2}.
	\end{align}	
	Thus proof is completed.

	\bibliographystyle{IEEEtran}
	\bibliography{IEEEabrv,Refference}
	
	\begin{IEEEbiography}[{\includegraphics[width=1in,height=1.05in,clip,keepaspectratio]{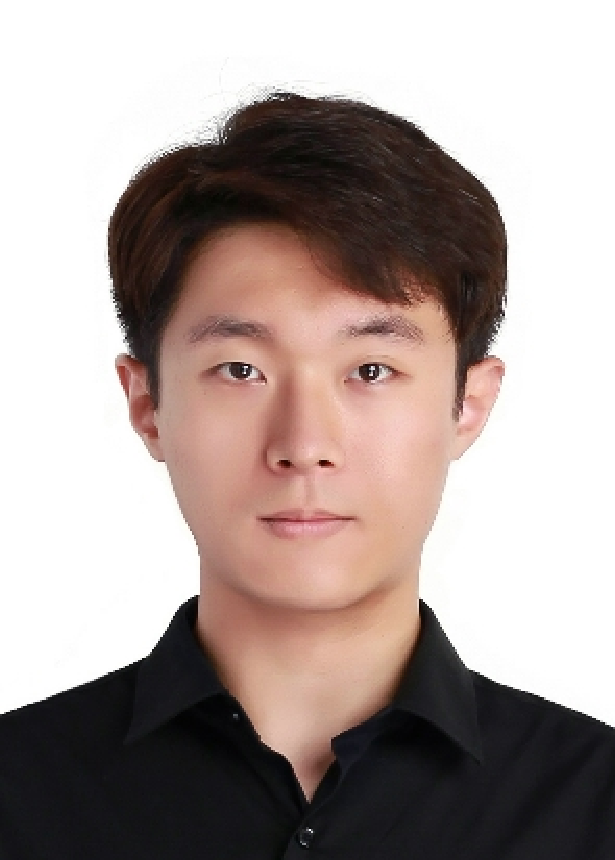}}]{Xiaochun Ge}
		received the B.S. degree in communication engineering from Beijing Institute of Technology, Beijing, China, in 2019, where he is currently pursuing the Ph.D. degree with the School of Information and Electronics. 	
		His main research interests include massive MIMO, intelligent reflecting surface, signal processing, and convex optimization.
	\end{IEEEbiography}

	\begin{IEEEbiography}[{\includegraphics[width=1in,height=1.05in,clip,keepaspectratio]{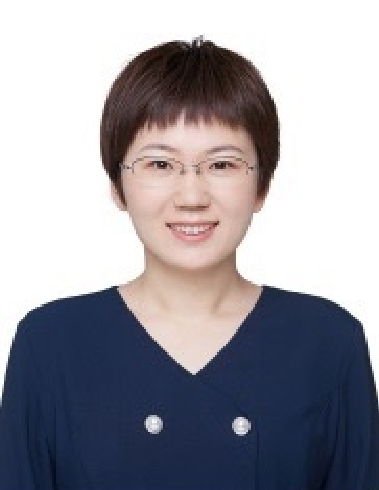}}]{Shanping Yu}
		received the Ph.D. degree from the Key Laboratory of Systems and Control, Academy of Mathematics and Systems Science, Chinese Academy of Sciences, Beijing, China, in 2017. She is currently an assistant professor with the School of Cyberspace Science and Technology, Beijing Institute of Technology, Beijing. Her research interests include social governance, intelligent computing, cyberspace security, and system optimization.
	\end{IEEEbiography}

	\begin{IEEEbiography}[{\includegraphics[width=1in,height=1.05in,clip,keepaspectratio]{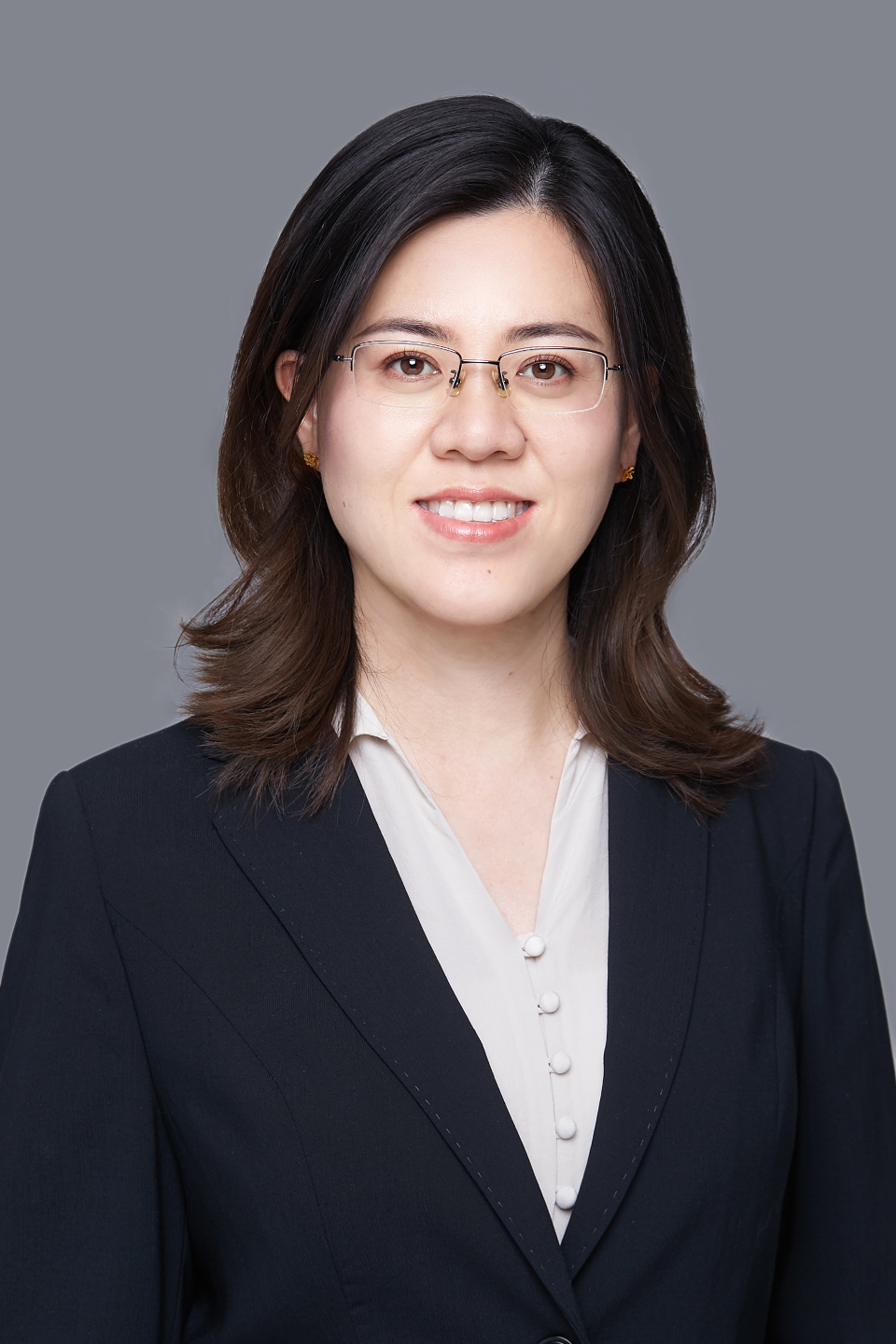}}]{Wenqian Shen}(Member, IEEE)
		received the B.S. degree from Xi'an Jiaotong University, Shaanxi, China, in 2013, and the Ph.D. degree from Tsinghua University, Beijing, China, in 2018. 
		She is currently an Associate Professor with the School of Information and Electronics, Beijing Institute of Technology, Beijing. 
		She has published several journal articles and conference papers in IEEE TRANSACTIONS ON SIGNAL PROCESSING, IEEE TRANSACTIONS ON COMMUNICATIONS, IEEE TRANSACTIONS ON VEHICULAR TECHNOLOGY, and IEEE ICC.
		Her research interests include massive MIMO and mmWave/THz communications. She received the IEEE Best Paper Award from the IEEE ICC 2017.
	\end{IEEEbiography}

	\begin{IEEEbiography}[{\includegraphics[width=1in,height=1.05in,clip,keepaspectratio]{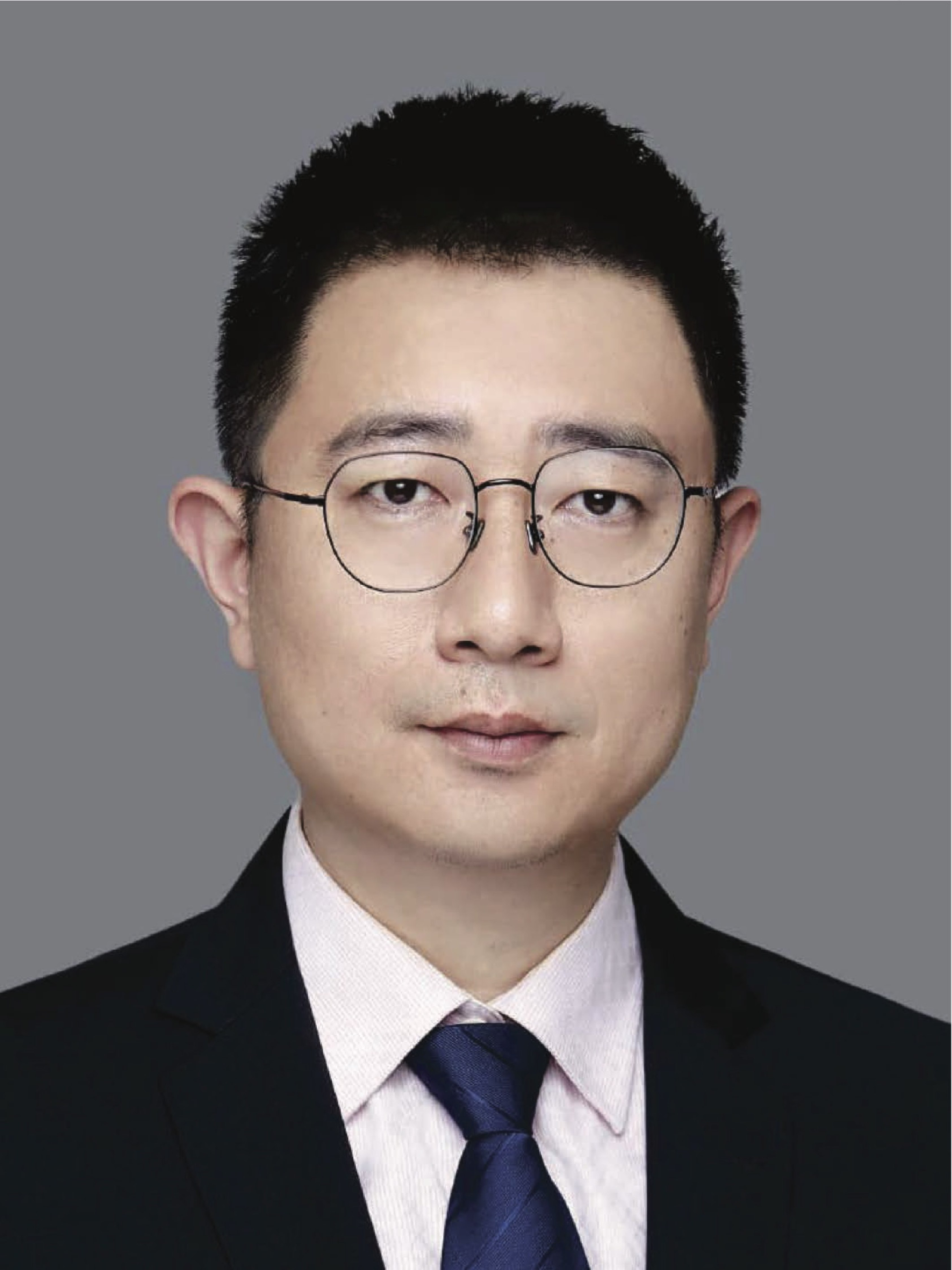}}]{Chengwen Xing}(Member, IEEE)
		received the B.Eng. degree from Xidian University, Xi'an, China, in 2005, and the Ph.D. degree from the University of Hong Kong, Hong Kong, China, in 2010. 
		Since September 2010, he has been with the School of Information and Electronics, Beijing Institute of Technology, Beijing, China, where he is currently a Full Professor. From September 2012 to December 2012, he was a Visiting Scholar at the University of Macau, Macau SAR, China. 
		His current research interests include machine learning, statistical signal processing, convex optimization, multivariate statistics, and array signal processing.
	\end{IEEEbiography}

	\begin{IEEEbiography}[{\includegraphics[width=1in,height=1.05in,clip,keepaspectratio]{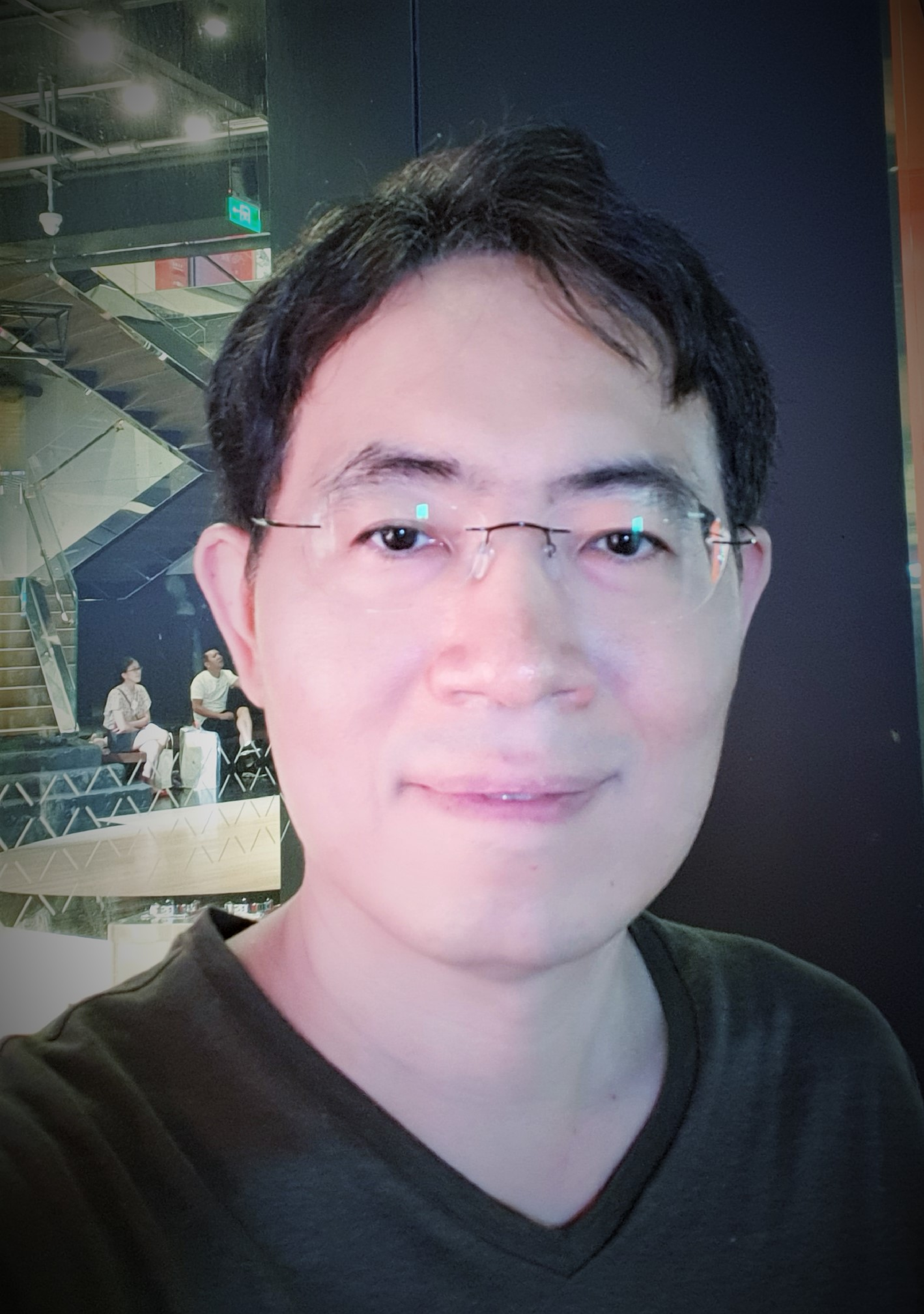}}]{Byonghyo Shim}(Senior Member, IEEE)
		received the B.S. and M.S. degrees in control and instrumentation engineering (currently electrical engineering) from Seoul National University (SNU), Seoul, South Korea, in 1995 and 1997, respectively, and the M.S. degree in mathematics and the Ph.D. degree in electrical and computer engineering from the University of Illinois at Urbana-Champaign (UIUC), Urbana, IL, USA, in 2004 and 2005, respectively. 
		From 1997 and 2000, he was with the Department of Electronics Engineering, Korean Air Force Academy, as an Officer (First Lieutenant) and an Academic Full-Time Instructor. 
		He had a short-time research position at LG Electronics, Texas Instruments, and Samsung Electronics, in 1997, 2004, and 2019, respectively. From 2005 to 2007, he was a Staff Engineer with Qualcomm Inc., San Diego, CA, USA. 
		From 2007 to 2014, he was an Associate Professor with the School of Information and Communication, Korea University, Seoul. Since September 2014, he has been with the Department of Electrical and Computer Engineering, SNU, where he is currently a Professor and the Director of the Institute of New Media and Communications. 
		His research interests include 5G and 6G wireless communications, statistical signal processing, deep learning, and information theory. He was a recipient of the M. E. Van Valkenburg Research Award from the ECE Department, University of Illinois, in 2005, the Hadong Young Engineer Award from IEIE in 2010, the Irwin Jacobs Award from Qualcomm and KICS in 2016, the Shinyang Research Award from the Engineering College of SNU in 2017, the Okawa Foundation Research Award in 2020, and the IEEE COMSOC AP Outstanding Paper Award in 2021.
		He was a Technical Committee Member of Signal Processing for Communications and Networking (SPCOM). 
		He has been serving as an Associate Editor for IEEE TRANSACTIONS ON WIRELESS COMMUNICATIONS (TWC), IEEE TRANSACTIONS ON SIGNAL PROCESSING (TSP), IEEE TRANSACTIONS ON COMMUNICATIONS (TCOM), IEEE TRANSACTIONS ON VEHICULAR TECHNOLOGY (TVT), IEEE WIRELESS COMMUNICATIONS LETTERS (WCL), and a Guest Editor for IEEE JOURNAL ON SELECTED AREAS IN COMMUNICATIONS (location awareness for radios and networks).
	\end{IEEEbiography}

\end{document}